\documentclass[12pt,preprint]{aastex}

\usepackage{color,amsmath,amsfonts,amssymb,amsthm}

\usepackage{mark_newcommands}
\usepackage{aas_macros}  
\usepackage{mark_journals}

\begin{document} 
\title{Baroclinic Vorticity Production in Protoplanetary Disks\\Part I: Vortex Formation}
\author{Mark R. Petersen}
\affil{Dept. of Applied Mathematics, University of Colorado at Boulder\\present address: Los Alamos National Laboratory\\  Computer and Computational Science Div. and Center for Nonlinear Studies}
\email{mpetersen@lanl.gov}
\author{Keith Julien}
\affil{Dept. of Applied Mathematics, University of Colorado at Boulder}
\author{Glen R. Stewart}
\affil{Laboratory for Atmospheric and Space Physics, University of Colorado at Boulder}


\begin{abstract}
The formation of vortices in protoplanetary disks is explored via pseudo-spectral numerical simulations of an anelastic-gas model.  This model is a coupled set of equations for vorticity and temperature in two dimensions which includes baroclinic vorticity production and radiative cooling.  Vortex formation is unambiguously shown to be caused by baroclinicity because (1) these simulations have zero initial perturbation vorticity and a nonzero initial temperature distribution; and (2) turning off the baroclinic term halts vortex formation, as shown by an immediate drop in kinetic energy and vorticity.  
Vortex strength increases with: larger background temperature gradients; warmer background temperatures; larger initial temperature perturbations; higher Reynolds number; and higher resolution.
In the simulations presented here vortices form when the background temperatures are $\sim 200K$ and vary radially as $r^{-0.25}$, the initial vorticity perturbations are zero, the initial temperature perturbations are 5\% of the background, and the Reynolds number is $10^9$.  
A sensitivity study consisting of 74 simulations showed that as resolution and Reynolds number increase, vortices can form with smaller initial temperature perturbations, lower background temperatures, and smaller background temperature gradients.
For the parameter ranges of these simulations, the disk is shown to be convectively stable by the Solberg-H{\o}iland criteria.

\end{abstract}




\section{Introduction \label{s_introduction}}

Planetary formation theories come in two flavors: core accretion, where particles build from millimeter-sized dust grains to kilometer-sized planetesimals and eventually to planetary cores \citep{Wetherill90areps}; and gravitational instability, where part of a massive disk quickly collapses to form giant planets by gravitational self-attraction \citep{Boss97sc}.  Each theory has its own difficulties.  Core accretion is an extremely slow process---it is thought that core accretion would take 8 to 48 million years to build a Jupiter-sized planetary core \citep{Pollack_ea96ic}.  This is inconsistent with the estimated disk life-time of a few million years \citep{Briceno_ea01sc,Bally_ea98apj}.  Also, the mechanism for the cohesion of meter-sized particles is poorly understood; they are too small to be influenced by gravity, but too large for agglomeration by collisional coagulation in a turbulent gas disk \citep{Weidenschilling84ic,Weidenschilling_Cuzzi93proc}.  Gravitational instabilities could form planets fast enough, but the densities required for instabilities are hard to attain: high gas densities throughout a thick disk would produce global gravitational instabilities that would transport a substantial fraction of the mass into the central star; a thin, high density layer of solids in the midplane of the disk may be precluded by shear turbulence in the vertical \citep{Cuzzi93ic}.

Protoplanetary disks initially consist of about 99\% gas and 1\% dust \citep{Hayashi81ptps}.  Although most discussions of planetary formation only consider solid materials, the gaseous dynamics can have a large influence on the distribution of the solids and thus the subsequent planet building.  Vortices in the gas disk are extremely efficient at capturing meter-size particles \citep{Tanga_ea96ic, Johansen_ea04aa, Klahr_Bodenheimer06apj}, so particle concentrations inside vortices would be several orders of magnitude higher than in the surrounding gas \citep{Barge_Sommeria95aa}.  Thus vortices would be beneficial for either theory: in core accretion, the rate of accretion depends strongly of the local concentration of particles in the disk \citep{Pollack_ea96ic}; similarly, a high local density of solids within a vortex could initiate a gravitational instability without requiring a high density throughout the disk.  The disks considered here are cool enough that they are only weakly ionized, and therefore turbulence due to magnetohydrodynamic effects is not considered.

In order to know whether vortices play a role in planetary formation, we must understand what conditions are required for vortex formation and longevity.  Several such models have been introduced in the literature in the last decade.
\citet{Bracco_ea99pf} used a two-dimensional incompressible shallow water model with a background Keplerian rotation to study the inverse energy cascade.  They found that if the initial energy of the disturbance is less than about $10^{-3}$ of the energy in the Keplerian flow, the vorticity fluctuations shear away and the disk returns to its unperturbed velocity profile.  For larger initial energies, anticyclonic vortices form and merge, while cyclonic vortices shear out.    \citet{Godon_Livio99apj} studied vortex longevity with a compressible Navier-Stokes pseudospectral model.  Vortex life-time, as measured by the maximum vorticity, was highly dependent on the Reynolds number; at $Re=10^4$--$10^5$ the maximum vorticity had an e-folding time of 60 orbital periods.  They also found that Keplerian shear inhibits the formation of vortices larger than the scale length $L_s=\sqrt{v/(\p_r\Omega)}$, where $v$ is the rotational velocity of the vortex and $\Omega$ is the angular velocity of the disk.

The only source of vorticity in two-dimensional (2D) models is baroclinicity, which appears as $\del p \times \del \rho$ in the vorticity equation, where $p$ is the pressure and $\rho$ is the density.  In the atmosphere, baroclinicity usually operates in the vertical and produces effects such as land-sea breezes when density and pressure (or temperature) gradients do not align.  In the geometry of a protoplanetary disk, density stratification is in the radial, so azimuthal variations in pressure will produce vorticity.  Most numerical models of protoplanetary disks do not include the effects of baroclinicity and can therefore only simulate decaying turbulence from some initial vorticity distribution.  For example, the incompressible fluid models used by \citet{Bracco_ea99pf} and \citet{Umurhan_Regev04aa} preclude any thermodynamic effects;  \citet{Godon_Livio99apj} and \citet{Johnson_Gammie05apjB} both assume a polytropic relation $p=K\rho^\gamma$ so that $\del p \times \del \rho=0$ and consequently baroclinic effects are not present.  

Vortex formation {\it due to baroclinic effects} was studied by \citet{Klahr_Bodenheimer03apj}, who found that the generation of turbulence and vortices depends strongly on the radial temperature gradient.  They observed vortices in both three-dimensional (3D) simulations with a full radiative scheme (a flux-limited radiative transport with a simple dust opacity), and 2D simulations where the radiative transport scheme was replaced by a specified background temperature distribution.  When the radial temperature was constant, no turbulence was produced; when $T\sim r^{-1}$, the flow became turbulent and vortices formed.  The results of \citet{Klahr_Bodenheimer03apj} are the motivation for the current study; we aim to elucidate how baroclinic vorticity production depends on background temperature, background temperature gradient, and the thermal diffusion.  Both our work and the simulations in \citet{Klahr_Bodenheimer03apj} have disks which are {\it convectively stable} in the radial, based on the Solberg-H{\o}iland criteria.

The study of 3D vortices by \citet{Barranco_Marcus05apj} models an anelastic fluid using the Euler equation and the ideal gas law; thus baroclinic effects are included.  They found that vortices that extend several scale heights vertically are unstable, as are short vortices centered at the midplane.  However, breaking internal gravity waves would spontaneously create 3D vortices centered 1--3 scale heights above or below the mid-plane.  This has obvious implications for the stability of the vortices that we observe in our 2D model, which are similar to their tall unstable vortices.  This issue is discussed further in the conclusion.  Notably, even though \citet{Barranco_Marcus05apj} include baroclinic effects, they do not have a radial temperature gradient in their model, so their vortices are not produced by the same process as \citet{Klahr_Bodenheimer03apj} or ours, which both require a nonzero radial temperature gradient.

This work is presented in two parts, where Part I studies vortex formation with small domain simulations for five orbital periods, and Part II studies vortex growth and longevity using a larger domain for hundreds of orbital periods.  This paper begins by describing the 2D anelastic gas equation set in \S \ref{s_cuppd_equations} and the pseudo-spectral numerical model in \S \ref{s_ppd_numerical_model}.  The results in \S \ref{s_results} include a description of the early evolution of these simulations, and a sensitivity analysis of how vortex formation depends on numerous parameters.  Conclusions follow in \S \ref{s_conclusions}.  The appendix includes details of the numerical model, including enforcement of stress-free boundary conditions with the influence matrix technique.  For conciseness there is little repetition between Parts I and II, so the reader is advised to read both together.

\section{Description of the Equation Set \label{s_cuppd_equations}}

Any numerical model of an astronomical object must make compromises between the
competing demands of including the essential physical processes, achieving
adequate numerical resolution, and exploring the relevant range of parameters.
Given the controversy generated by the large scale simulations published by
\citet{Klahr_Bodenheimer03apj}, we believe it is prudent to use a simplified model
that can help elucidate the necessary requirements for baroclinic vorticity
production in the 2D disk geometry.  More realistic models can be used to verify
the relevance of our model results to actual protoplanetary disks.   

In this
study we model a 2D anelastic gas, rather than compressible gas, in
order to make a fast numerical model that can explore a large swath of parameter
space by performing numerous simulations.  Anelastic models filter out the
acoustic waves present in compressible gases; the lack of these fast-propagating
acoustic waves relaxes the time step restrictions of the numerical modes,
thereby reducing the computational cost.  By using a fast numerical model, we
are also able to run simulations at higher resolutions than most published
models of astrophysical disks and thereby minimize the effects of numerical
viscosity.  Compressible disk simulations typically have such large numerical
viscosities that they can suppress the baroclinic production of vorticity we
want to explore \citep{Klahr_Bodenheimer03apj}.  A fully 3D anelastic model would
also be very expensive to run at the same resolution that we can easily explore
with our 2D model.  Finally, our model retains the essential physics of
interest, namely the baroclinic instability and radiative cooling.


To arrive at the anelastic limit, assume that thermodynamic perturbations are small compared to the background state and that radial displacements are significantly less that the length scale for radial gradients in the unperturbed surface density of the disk \citep{Bannon96jas}.  Then the conservation of mass equation,
\bea
\frac{D\rho}{Dt} + \rho \del\cdot \bu = 0, \label{cons_mass}
\eea
simplifies to the anelastic equation,
\bea
\del\cdot \left( \rho \bu \right) = 0,
\eea
where the density $\rho$ is constant in time and varies only in the direction of stratification.  Here $\bu$ is the velocity vector and $D/Dt$ is the material derivative.

The model equation set is inspired by \citet{Bannon96jas} and \citet{Scinocca_Shepherd92jas}, which are anelastic models of the atmosphere derived from conservation of momentum, conservation of mass, the second law of thermodynamics, and the ideal gas law.  These equations were modified to the geometry of a disk, where the background density and temperature stratification are in the radial and shear due to near-Keplerian rotation is in the azimuthal.  Assuming that the disk is thin and hydrostatic in the vertical leads to a 2D model.  A vorticity-stream function formulation is used so that only two prognostic variables, the vorticity and the potential temperature, must be computed by the numerical model.  

The model equations are
\bea
\zeta &=& \ds \frac{1}{r} \frac{\p}{\p r}
   \left( \frac{r}{\Sigma_0} \frac{\p \Psi}{\p r} \right)
   + \frac{1}{r^2 \Sigma_0} \frac{\p^2\Psi}{\p \phi^2} 
\label{an_stream_fxn}\\
\ds \frac{\p\zeta}{\p t}
  + \p\left(\Psi,\frac{1}{\Sigma_0}\zeta\right) \ds
  &=& \ds  \frac{c_p}{r}
      \dd{\pi_0}{r}\dd{\theta'}{\phi} 
\label{an_vorticity} \\
\ds \frac{\p \theta}{\p t}
  + \frac{1}{\Sigma_0}\p\left(\Psi,\theta\right)
  &=& \ds -\frac{\theta'}{\tau}.
\label{an_temperature}
\eea
The vertical component of vorticity, the stream function, and the potential temperature are the sum of a background and perturbation variables,
\beas
\zeta &=& \zeta_0(r) + \zeta'(r,\phi,t) \\
\Psi  &=& \Psi_0(r) +  \Psi'(r,\phi,t) \\
\theta  &=& \theta_0(r) + \theta'(r,\phi,t),
\eeas
where the primed perturbation quantities are small relative to the background.  Radial and azimuthal velocities $\bu=(u,v)$ are related to the streamfunction by $\Sigma_0 \bu = -\del\times\Psi {\bf \hat{z}}$.  Other variables include the surface density $\Sigma_0(r)$, Exner pressure $\pi_0(r)$, radiative cooling time $\tau$, specific heat at constant pressure $c_p$, time $t$, polar coordinates $(r,\phi)$, vertical unit vector ${\bf \hat{z}}$, and the Jacobian $\p(a,b)=(\p_ra\p_\phi b - \p_\phi a\p_rb)/r$.  The potential temperature and Exner pressure can be written in terms of the pressure $p$ and temperature $T$ as
\bea
\theta=T\left(\frac{p_0(r_{in})}{p}\right)^{R/c_p}, \;\;
\pi = \frac{T}{\theta} = \left(\frac{p}{p_0(r_{in})}\right)^{R/c_p},\nn
\eea
where $p_0(r_{in})$ is a reference pressure, $r_{in}$ is the inner radius of the annulus, and $R$ is the gas constant.  The background potential temperature profile, $\theta_0(r)$, is assumed to be maintained by a balance between radiative heating by the central star and radiative cooling to space.

The background variables obey a static background balance between the centrifugal force, gravity, and the pressure gradient,
\bea
\Omega_0^2r = \frac{\p \Phi}{\p r} + \frac{1}{\Sigma_0}\frac{\p p_0}{\p r},
\label{r_F_balance}
\eea
where $\Omega_0$ is the background angular velocity, $\Phi$ is the gravitational potential, and $p_0$ is the pressure integrated over the thickness of the disk.  This leading order balance has already been subtracted out of the vorticity equation (\ref{an_vorticity}).

Equation (\ref{an_vorticity}) is a conservation equation for potential vorticity, $\zeta/\Sigma_0$, where the baroclinic term on the right-hand side is a source of vorticity and the cause of the baroclinic instability that is central to our study.  This baroclinic term will look unfamiliar to most readers due to the variables and geometry used.  Recall that the common form comes from taking the cross-product of the pressure gradient term in the momentum equation,
\bea
\del\times\left(-\frac{1}{\rho} \del p \right)
 = \frac{1}{\rho^2} \del\rho\times\del p.
\eea
The equivalent operation in our case is
\bea
\del\times\left(-c_p \theta'\del\pi_0 \right)
 =-\frac{1}{r}\frac{\p}{\p\phi}\left(-c_p \theta'\frac{d\pi_0}{dr} \right)
 = \frac{c_p}{r}\frac{\p\theta'}{\p\phi} \frac{d\pi_0}{dr}.
\eea

In the numerical model, effective eddy dissipation terms are added to the vorticity and temperature equations for numerical stability, and many of the background variables drop out:
\bea
\zeta' &=& \ds \frac{1}{r} \frac{\p}{\p r}
   \left( \frac{r}{\Sigma_0} \frac{\p \Psi'}{\p r} \right)
   + \frac{1}{r^2 \Sigma_0} \frac{\p^2\Psi'}{\p \phi^2} 
\label{num_stream_fxn}\\
\ds \frac{\p\zeta'}{\p t}
  + \p\left(\Psi,\frac{1}{\Sigma_0}\zeta\right) \ds
  &=& \ds  \frac{c_p}{r}
      \dd{\pi_0}{r}\dd{\theta'}{\phi} 
  + \nu_e \del^2 \zeta' 
\label{num_vorticity} \\
\ds \frac{\p \theta'}{\p t}
  + \frac{1}{\Sigma_0}\p\left(\Psi,\theta\right)
  &=& \ds -\frac{\theta'}{\tau} + \kappa_e \del^2 \theta'.
\label{num_temperature}
\eea
Here $\nu_e$ is the viscosity and $\kappa_e$ is the thermal dissipation.

The equation set is then nondimensionalized using the length scale $L_{sc}=r_{mid}$; the time scale $t_{sc}=2\pi/\Omega_0(r_{mid})$, which is one orbital period; and typical temperature and density scales.  For the standard domain, $L_{sc}=7.5$AU and $t_{sc}=20$ (earth) years, background temperatures are 125--600K, and background surface densities are 350--1000g/cm$^{-2}$.  These scales let us define the Reynolds and Peclet numbers based on the eddy dissipation terms, 
\beas
Re = \frac{L_{sc}^2}{\nu_e t_{sc}}, \hspace{1cm}
Pe = \frac{L_{sc}^2}{\kappa_e t_{sc}},
\eeas
which measure the ratio of momentum advection to kinetic energy dissipation and thermal advection to thermal dissipation, respectively.  For a particular grid spacing, $Re$ and $Pe$ must be small enough to avoid building up energy at the smallest scales.  Simulations with a larger number of gridpoints allow larger values of $Re$ and $Pe$, which reduces dissipation so that coherent structures are more long-lived.  For brevity, the nondimensionalized equations are not presented here.

Factors of the background surface density $\Sigma_0$ appear in the model equations because $\Psi$ is a momentum stream function, defined as $\Sigma_0 \bu = -\del\times\Psi {\bf \hat{z}}$ where $\bu=(u,v)$, the radial and azimuthal velocities, and ${\bf \hat{z}}$ is a vertical unit vector.  A momentum stream function was chosen because the anelastic equation, $\del\cdot(\Sigma_0\bu)=0$ dictates that the product $\Sigma_0\bu$ is divergence free.  (In incompressible flows, $\del\cdot\bu=0$ and the stream function is simply $\bu = -\del\times\Psi{\bf \hat{z}}$.)  Equation (\ref{num_stream_fxn}) is nearly a polar Laplacian $\zeta'=\del^2\Psi'$.  The density factors present an extra challenge for the inversion step to find $\Psi'$ (appendix \ref{a_stream_function}).  The Jacobians on the left-hand side of (\ref{num_vorticity}) and (\ref{num_temperature}) represent a source in the Eulerian frame due to advection, and combine with the time derivatives to form material derivatives in polar coordinates.


Equation (\ref{num_temperature}) is an advection-diffusion equation for potential temperature perturbation, $\theta'$.  The two terms on the right-hand side are both decay terms: the Laplacian operator $\del^2$ imparts an effective eddy diffusion which dissipates more quickly at the smallest scales; the radiative damping term, $-\theta'/\tau$, is a simplified model of blackbody radiation and dissipates perturbations equally at all scales so as to relax the temperature field back to the background value.  The background potential temperature profile, $\theta_0(r)$, is assumed to be maintained by a balance between solar heating and radiative cooling to space.

The temperature equation (\ref{num_temperature}) is coupled to the vorticity only by its advection (Jacobian) term, while the vorticity equation (\ref{num_vorticity}) is coupled to temperature by its baroclinic term.  This coupling allows the baroclinic feedback between vorticity and temperature perturbations to occur, which can cause nonlinear growth leading to turbulence.  Understanding what parameters affect the strength of this baroclinic feedback is one of the main goals of this work.

\section{The Numerical Model\label{s_ppd_numerical_model}}
A two-dimensional spectral model was created to investigate the nonlinear behavior of the equation set.  Spectral methods have the advantages of exponentially convergent spatial representation and efficient inversion of derivative operators.  Vorticity and temperature fields on the annulus are represented using Fourier basis functions in the azimuthal and Chebyshev basis functions in the radial.  The time-stepping algorithm is a third-order semi-implicit Runge-Kutta routine which is optimized for efficient memory use \citep{Spalart_ea91jcp}.  Other authors that have created spectral models on an annulus or disk using Fourier-Chebyshev basis functions include \citet{Godon97apj}, \citet{Bergeron00jfm}, and \citet{Torres_Coursias99SIAMjsc}.

The Laplacian operator, which is inverted at each stage of the semi-implicit Runge-Kutta routine, is an upper triangular matrix when applied to the Chebyshev coefficients for each Fourier mode.  This upper triangular matrix was reduced algebraically to a banded diagonal of width six, greatly improving the efficiency of the inversion step, which is potentially a very time consuming operation (see appendix \ref{a_inversion}).  Specifically, inverting the Laplacian operator with $N$ Chebyshev modes and $M$ Fourier modes, the operation count is $\mathcal{O}(MN^2)$ for a back-solve routine on a triangular matrix, and is reduced to $\mathcal{O}(MN)$ for a banded diagonal.

A standard polar Laplacian operator is inverted for the vorticity and temperature equations, (\ref{num_vorticity}) and (\ref{num_temperature}).  For the inversion of the near-Laplacian with surface density factors in (\ref{num_stream_fxn}), it is not possible to solve for $\Psi$ using an arbitrary function $\Sigma_0(r)$.  However, a simple power function $\Sigma_0 = cr^d$ only adds a constant in front of one of the radial derivative terms, as described in the appendix \ref{a_stream_function}.  This method allowed us to incorporate radial variations in the surface density.  The restriction to power laws is not overly prohibitive, as they are standard test cases in other studies \citep{Klahr_Bodenheimer03apj, Godon_Livio99apj}.

All derivatives are computed in spectral space using the Fourier-Chebyshev coefficients.  The products that appear in the advective and baroclinic terms are computed at each grid point in physical space.  In this pseudospectral method, the vorticity and temperature fields must be transformed between physical and spectral space at every step.  We use the FFTW fast Fourier transform package \citep{Frigo_Johnson98proc},  which has been benchmarked as one of the fastest FFT algorithms on numerous platforms.

The first task in running fluid simulations is to find the appropriate dissipation rate for the grid spacing.  The goal is to have as little dissipation as possible (i.e., large Reynolds and Peclet numbers), while dissipating enough energy at the smallest scales to prevent grid-scale oscillations and numerical instabilities.  Calculating a one-dimensional energy spectrum in Fourier-Chebyshev space is not straightforward, so separate plots of energy versus Fourier modes and Chebyshev modes were checked to ensure that the energy decays properly at small scales.

Additional viscosity was added at large radial wavenumbers to dissipate energy near the gridscale, so that the radial Laplacian terms in (\ref{num_vorticity}) and (\ref{num_temperature}) have the coefficient
\bea
\nu_{coef}=\frac{1}{2}\left[\nu_f+1+\left(\nu_f -1\right)
\tanh\left(\frac{k-k_c}{k_w}\right)  \right]
\label{visc_factor}
\eea
in spectral space. For wavenumbers somewhat smaller than the critical wavenumber $k_c$,  $\nu_{coef}\sim1$ and the diffusion operators are unaffected; for wavenumbers $k>>k_c$ the coefficients $1/Re$ and $1/Pe$ are multiplied by the factor $\nu_{coef}\sim\nu_f$.  Typical simulations used a factor $\nu_f$ of 500, a critical wavenumber $k_c$ of four-fifths the maximum wavenumber, and a transition width $k_w$ of 10.  These parameters were chosen so that only the largest Chebyshev (radial) basis functions experience additional damping.  It was found that hyperviscosity was not needed for the Fourier (azimuthal) basis functions.  With the addition of this hyperviscosity, we were able to increase the effective Reynolds number from $5\times10^4$ to $4\times10^7$, and develop much finer-scale structures at the same resolution.  Including hyperviscosity does not affect the conclusions of this study; all the simulations in Table \ref{t_parameters} were also run without hyperviscosity and produced the same qualitative results.

The azimuthal domain can be chosen from a full annulus or any fractional annulus $2\pi/2^n$, such as a quarter or 64th.  Fourier basis functions are periodic in the azimuthal in all of these cases, but the indexing of the basis functions and their derivatives must be handled with care.  For example, if a function is represented on the full annulus as 
\bea
\sum_{k=0}^N a_k e^{ik\phi}
\eea
then on the half annulus all of the odd modes must be zero, so only the even modes $a_0,a_2,a_4,a_6\ldots$ survive.  For efficient memory storage, only the nonzero coefficients are kept in the array, so the multiplication factor of a derivative $d/d\phi \sim ik$ is not the same as the array index.  The simulations in this paper are on a 64th annulus, with a resolution of $128\times 128$ or $256\times 256$.  Paper II presents simulations on the quarter annulus with resolutions of  $256\times 256$ and $512\times 512$.

\subsection{Boundary conditions}
Fourier basis functions are the natural choice for the periodic boundary conditions in the azimuthal direction of the annulus.  The radial perturbation temperature boundary condition is zero flux (Neumann), $\p_r \theta'=0$, at both the inner and outer boundaries.  Physically, the zero flux conditions means that the disk is not influenced by heating from outside of the computational domain.  They are enforced during the inversion of the Laplacian operator using the tau method, where two rows of the Laplacian matrix are substituted with the boundary constraints.  This effectively means that the projection of the temperature perturbation variable onto the highest two Chebyshev polynomial modes are exchanged for constraints on the boundary conditions \citep{Fornberg96bk}.

The dynamic boundary conditions are stress-free and preclude flow normal to the inner and outer boundary,
\bea
\frac{\p v'}{\p r} - \frac{v'}{r}=0,
\;\;\;u'=0.
\eea
This choice is sensible for astrophysical applications, as there are no solid surfaces to impart stress at the boundaries.  Inflow/outflow conditions that drive  the perturbation flow through boundary forcing are not considered in this study.  The stress-free boundary conditions are rewritten in terms of the stream function and vorticity, and are enforced using the influence matrix technique (appendix \ref{a_stress_free_bc}).

\subsection{Initial Conditions\label{s_ic}}
Initial conditions used for the perturbation variables $\zeta'$ and $\theta'$ include: (1) Fourier-Chebyshev modes; (2) Fourier-Bessel modes; (3) Fourier-sine modes; (4) a collection of vortices; and (5) a specified energy spectrum which is random in phase.  Individual modes were used extensively in testing the code, as described in the model validation section \ref{s_model_validation}.  Shearing effects and orbital time periods are clearly visible with the Fourier-sine modes.  Individual vortices were used to compare the behavior of cyclonic versus anticyclonic vortices, and to observe vortex merger.

\begin{figure}[tbh]
 \center
 \scalebox{.12}{\includegraphics{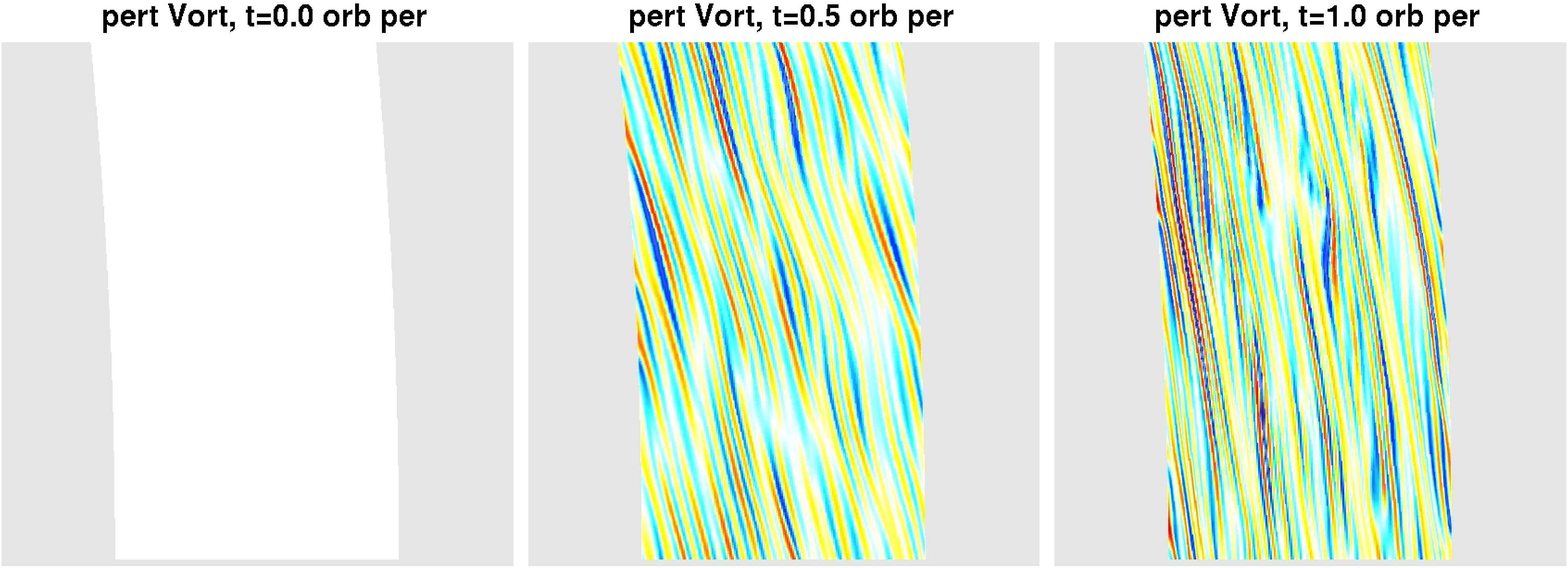}} 
 \scalebox{.4}{\includegraphics{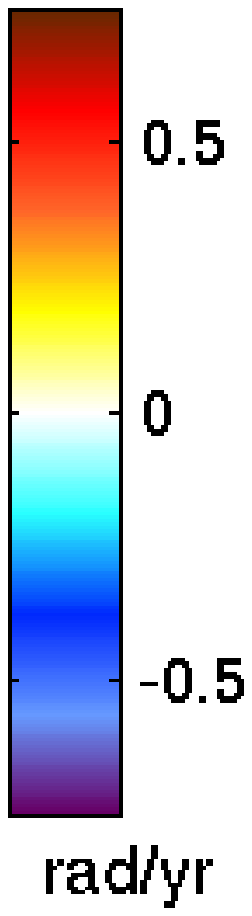}}\\ 
\scalebox{.12}{\includegraphics{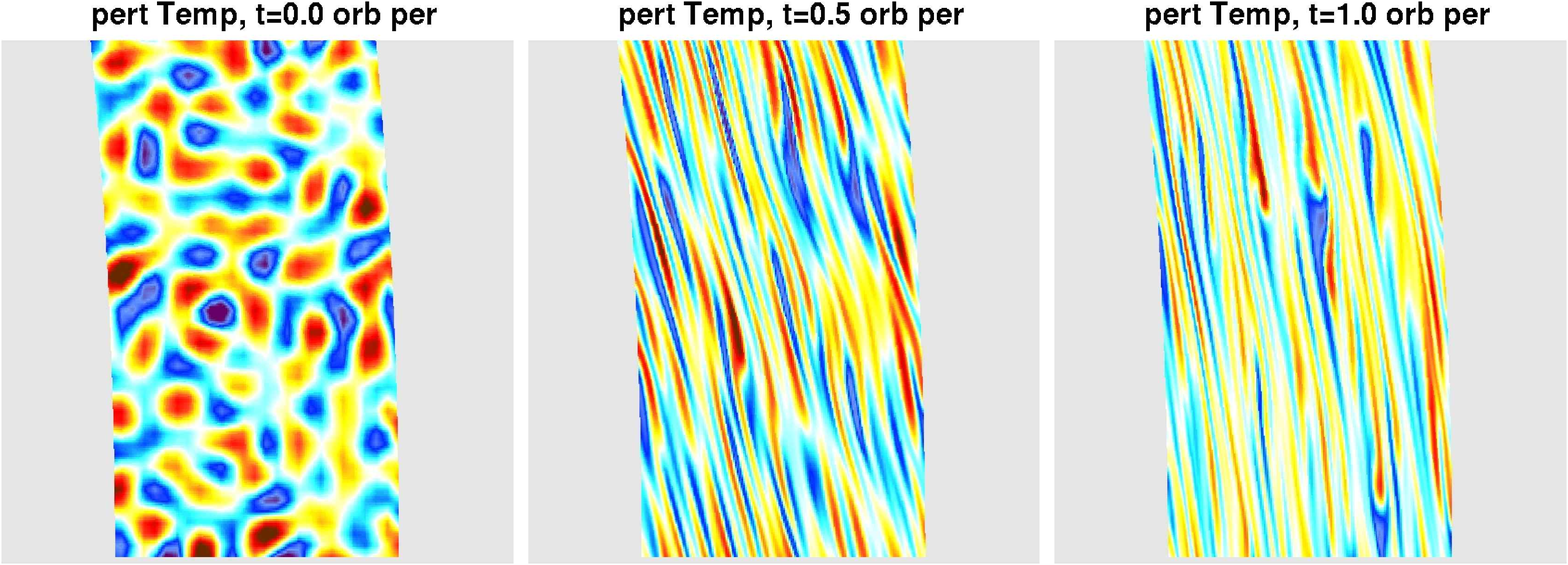}} 
\scalebox{.4}{\includegraphics{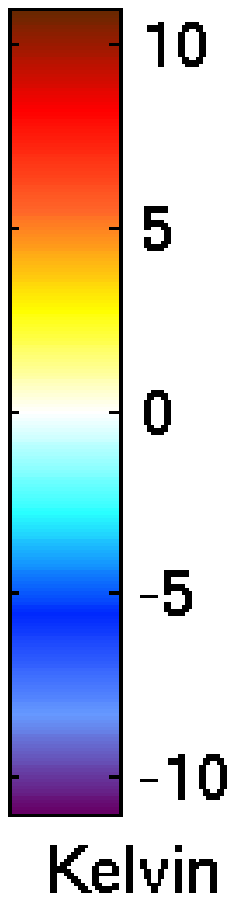}} 
 \caption{\label{f_early1} 
The perturbation vorticity $\zeta'$ (top) and perturbation temperature $T'$ (bottom) from simulation A, with a radial extent of 0.5 AU on a 1/64th annulus.  The time $t$ refers to the orbital period in the middle of the annulus.  All simulations begin with zero vorticity perturbation and a random initial temperature perturbation.  Vorticity is immediately produced by the baroclinic term, while the initial temperature perturbations shear out and decay.}
\end{figure}

The standard initial condition used in turbulence studies is a specified spectrum which is random in phase \citep{McWilliams90jfm_a,Bracco_ea00pf}.  A separate Cartesian periodic model was used to create these initial vorticity or temperature fields.  The initial energy spectrum is
$E(k) \sim {k^\alpha}/({k^{2\alpha}+k_0^{2\alpha}})$,
 where $k$ is the Cartesian wave number modulus and $k_0$ and $\alpha$ are parameters that change the shape of the spectrum.  The simulations discussed in this paper use $k_0=10$ and $\alpha=10$, as shown in the initial temperature perturbation in Fig. \ref{f_early1}.  Sensitivity studies with showed that the qualitative characteristics of the simulations did not depend on the particular value of $k_0$.  

High resolution initial conditions from the Cartesian model were interpolated to the Fourier-Chebyshev gridpoints of the annulus using bilinear interpolation.  The inner and outer radial edges were interpolated to zero to conform with boundary conditions, and the azimuthal edge was linearly interpolated to a periodic azimuthal boundary.

\subsection{Model validation \label{s_model_validation}}
The numerical model was validated by direct comparison with exact linear solutions using Fourier-Bessel functions.  In the absence of the nonlinear and baroclinic terms, (\ref{num_vorticity}) and (\ref{num_temperature}) become heat equations in vorticity and temperature.  For a single Fourier-Bessel basis function of the same mode number,
\bea
\del^2 J_k(r)e^{ik\phi} = -J_k(r)e^{ik\phi},
\eea
where $J_k(r)$ is the $k^{th}$ order Bessel function of the first kind in the radial direction and $e^{ik\phi}$ is the $k^{th}$ Fourier mode in the azimuthal.  Thus the exact solutions of (\ref{num_vorticity}) and (\ref{num_temperature}) decay exponentially with $e$-folding times of $1/Re$ and $1/\tau + 1/Pe$, respectively.  Including the baroclinic term adds a non-homogeneous term to the exact solution of the temperature.  

As there are no exact solutions to the full nonlinear model, inclusion of the advection terms were tested in other ways.  The Jacobian subroutine was tested individually and compared to analytic expressions to verify proper operation.  When the advection is computed using only the background temperature, vorticity, and stream function, it becomes a linear term and an exact solution is available for comparison.  Finally, qualitative results of vortex co-rotation, merger and translation showed that the numerical model was working as expected.

\section{Results\label{s_results}}

Hundreds of simulations were performed to study the effects of varying the model parameters.  The simulations discussed in this paper, shown in Table \ref{t_parameters}, vary parameters that affect vortex formation.  The domain of these simulations have a radial and azimuthal extent of $r\in[9.5,10]$AU and $\phi\in[0,\pi/32]$, respectively, and last for five orbital periods (here an orbital period is a full orbit of $2\pi$ at mid-annulus, $r_{mid}=9.75$AU).  Paper II discusses simulations with larger domains and longer duration, up to 600 orbital periods.

\begin{figure}[tbh]
\center
\scalebox{.8}{\includegraphics{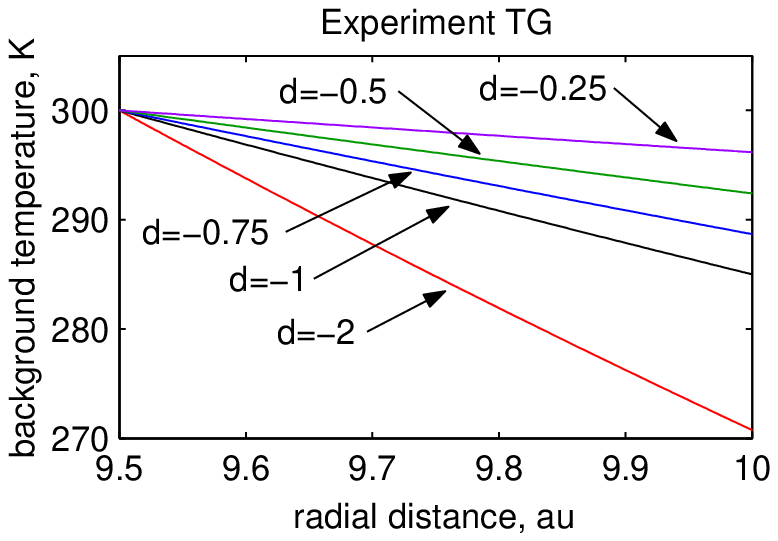}}
\scalebox{.8}{\includegraphics{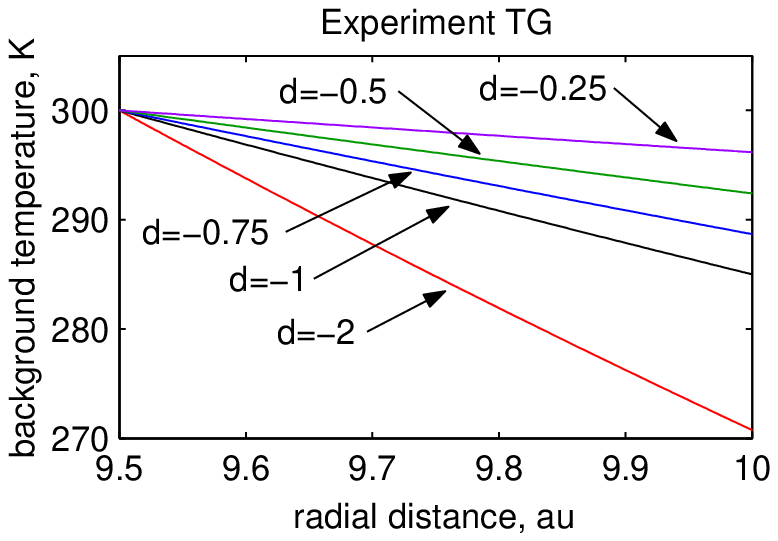}}\\ 
\scalebox{.8}{\includegraphics{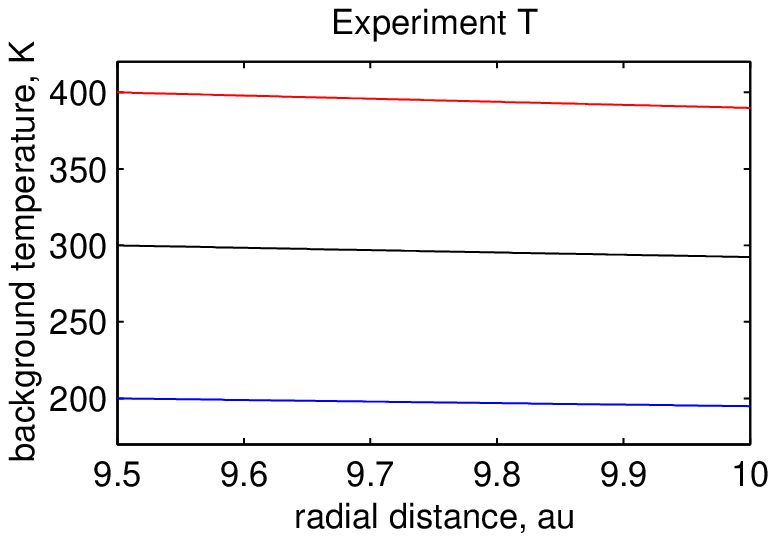}}
\scalebox{.8}{\includegraphics{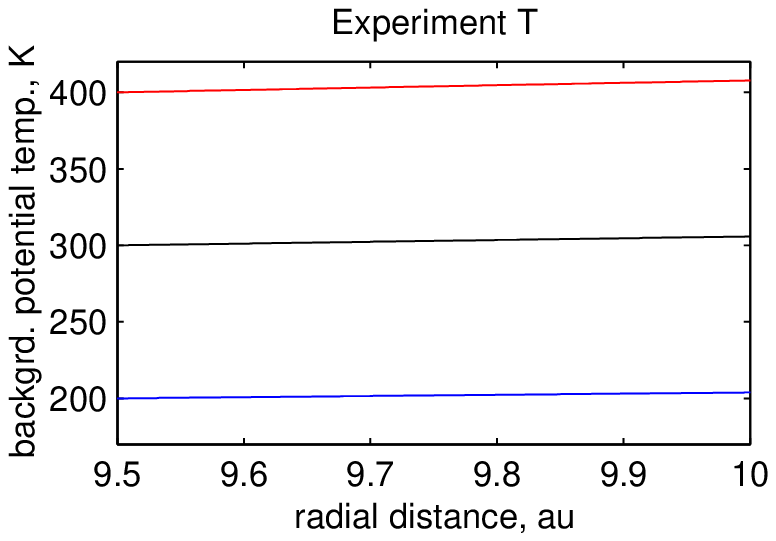}}
\caption{\label{f_bkrd_T} 
Background temperature $T_0$ (left) and potential temperature $\theta_0$ (right) profiles for simulations TG (top) and T (bottom), where $T_0(r) = c (r/9.5AU)^d$, and $d$ is shown.  Simulations A, B, and Tau use $d=-0.75$ and P uses $d=-0.5$.
}\end{figure}

The background surface density and temperature are power functions,
\bea
\Sigma_0(r) = a\left(\frac{r}{r_{in}}\right)^b, \;\;\; 
T_0(r) = c\left(\frac{r}{r_{in}}\right)^d,
\eea
where $r_{in}$ is the inner radius of the annulus.  All simulations presented here use $a=500$g cm$^{-2}$ and $b=-3/2$, while $c$ and $d$ are varied and shown in Table \ref{t_parameters}.  The resulting background temperature distributions and corresponding potential temperature distributions are shown in Fig. \ref{f_bkrd_T}.  All the remaining results in this paper are presented in terms of thermal temperature $T$, rather than potential temperature $\theta$.  This is a choice of convention; the model equations (\ref{an_temperature}) are in terms of potential temperature, but thermal temperature is more intuitive and easier to compare to observations.  

The Schwarzschild criterion states that a disk is convectively stable in the absence of rotation and shear if entropy, $S$, increases radially, i.e. $dS/dr>0$ \citep{Schwarzschild58bk}.  Entropy is related to potential temperature as 
\bea
S = c_p \ln\left(\frac{\theta}{\theta_{in}}\right)
\eea
where $\theta_{in}$ is a reference temperature, so the Schwarzschild criterion in terms of potential temperature is simply $d\theta/dr>0$.  The Schwarzschild criterion is satisfied for simulations where $d>-1$ (Fig. \ref{f_bkrd_T}).

In the presence of differential rotation, a sufficient condition for convective stability is the Solberg-H{\o}iland criterion,
\bea
\frac{1}{r^3}\dd{j^2}{r} - \frac{1}{c_p\rho}\del p \cdot \del S > 0,
\eea
where $j=r^2\Omega$ is angular momentum per unit mass \citep{Tassoul00bk,Ruediger_ea02aa}.  In our variables, this criterion is
\bea
\frac{1}{r^3}\dd{j^2}{r} 
  - \frac{1}{\theta_0\Sigma_0}\dd{R\Sigma_0 T_0}{r}\dd{\theta_0}{r} > 0.
\label{SH}
\eea
Differential rotation has a stabilizing effect on disks.  In other words, disks which do not satisfy the Schwarzschild criterion may still satisfy the Solberg-H{\o}iland criterion if the shear ($\p j^2/\p r$) is large enough.  This is true for the simulations presented here; when $d=-1$ and $d=-2$, the left-hand side of (\ref{SH}) ranges between 0.034 and 0.044 years$^{-2}$.  In summary, all of the simulations discussed in this paper are convectively stable.

\begin{figure}[tbh]
\center
\scalebox{.8}{\includegraphics{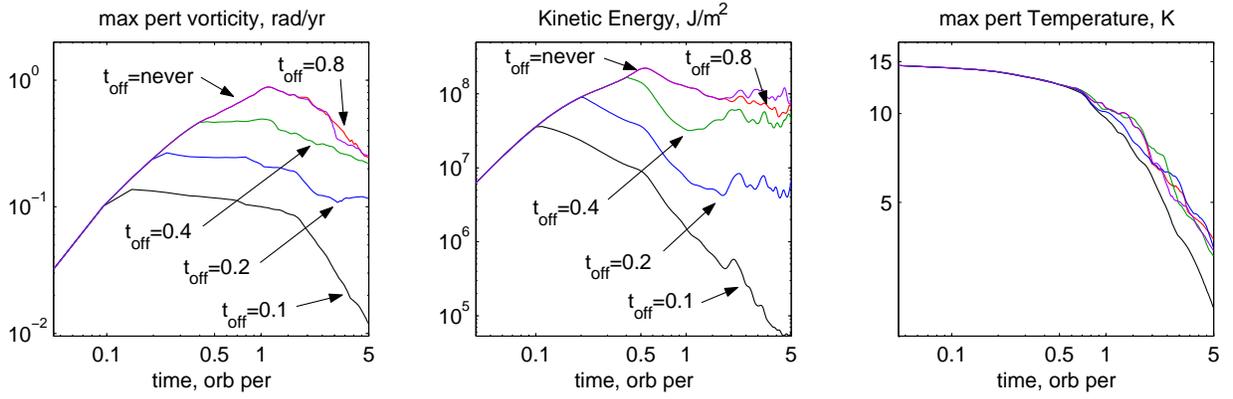}} 
\caption{\label{f_bcl_off_early} 
Comparison of maximum perturbation vorticity $|\zeta'|$ (left), perturbation kinetic energy (center), and maximum perturbation temperature $|T'|$ (right) for simulation series B, where the baroclinic term was turned off at the time indicated.  When the baroclinic term is turned off, there is no longer a source of vorticity.  This shows that the vortices are produced by the baroclinic instability.  The reference simulation ($t_{off}$=never) is simulation A.
}
\end{figure}

The initial condition for the perturbation temperature was a specified distribution in spectral space (Sect \ref{s_ic}) and a magnitude in physical space which is 5\% of the maximum background temperature.  The initial vorticity is zero; the only way for vorticity to be produced in these simulations is through the baroclinic term.  In a typical simulation, the initial temperature perturbations create vorticity perturbations in this way during the first several orbital periods (Fig. \ref{f_early1}).  The maximum perturbation vorticity $|\zeta'|$, the perturbation kinetic energy $|u'^2+v'^2|/2$, and the maximum perturbation temperature $|T'|$ for simulation A are shown in Fig. \ref{f_bcl_off_early}.  The vorticity and kinetic energy grow at early times due to the baroclinic term, while the temperature perturbations die off due to thermal dissipation.  After about one orbital period, the baroclinic vorticity production stops because the temperature perturbations are so low that $\p \theta'/\p \phi$ in the baroclinic term is also small.  
During the first orbital period, differential rotation causes the vorticity
perturbations to be stretched out into narrow bands of positive and negative
vorticity.  At the same time, the local radial gradient in the vorticity
perturbation rapidly grows to a magnitude that is comparable to the vorticity
gradient of the background Keplerian flow.   This kind of shear-induced
amplification of vorticity perturbations in astrophysical disks has been
described in many recent publications \citep{Chagelishvili_ea03aap,Tevzadze_ea03aap,Yecko04aap, Umurhan_Regev04aa}.  In particular, 
\citet{Afshordi_ea05apj}
 show that this transient amplification occurs in three-dimensional disks
so long as the radial wave number is large compared to both the azimuthal and
vertical wave numbers.

\begin{figure}[tbh]
 \center
 \scalebox{.12}{\includegraphics{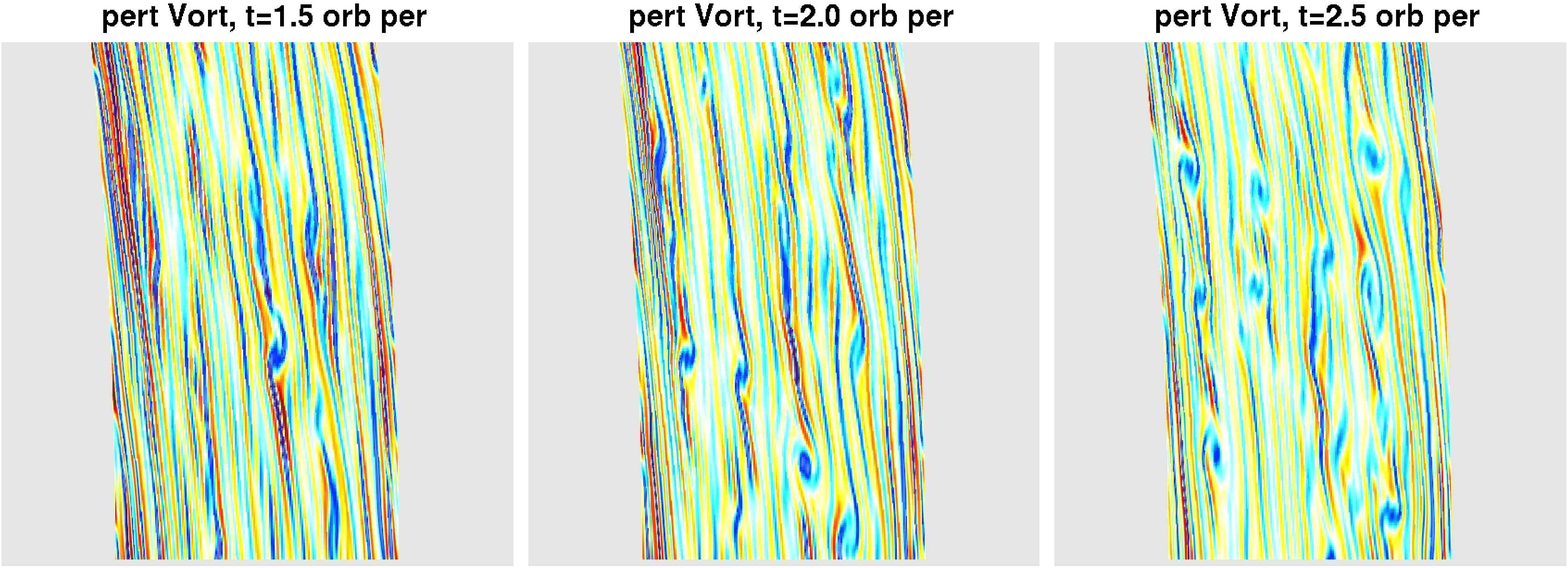}} 
 \scalebox{.4}{\includegraphics{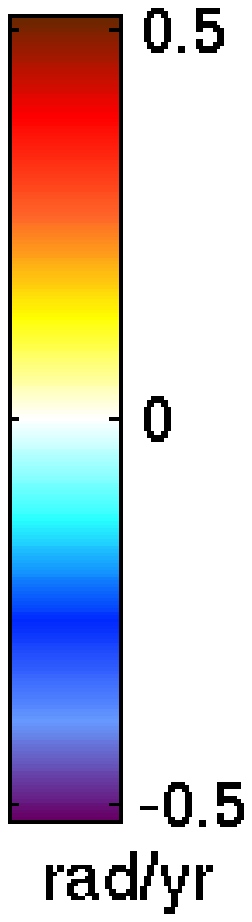}}\\ 
\scalebox{.12}{\includegraphics{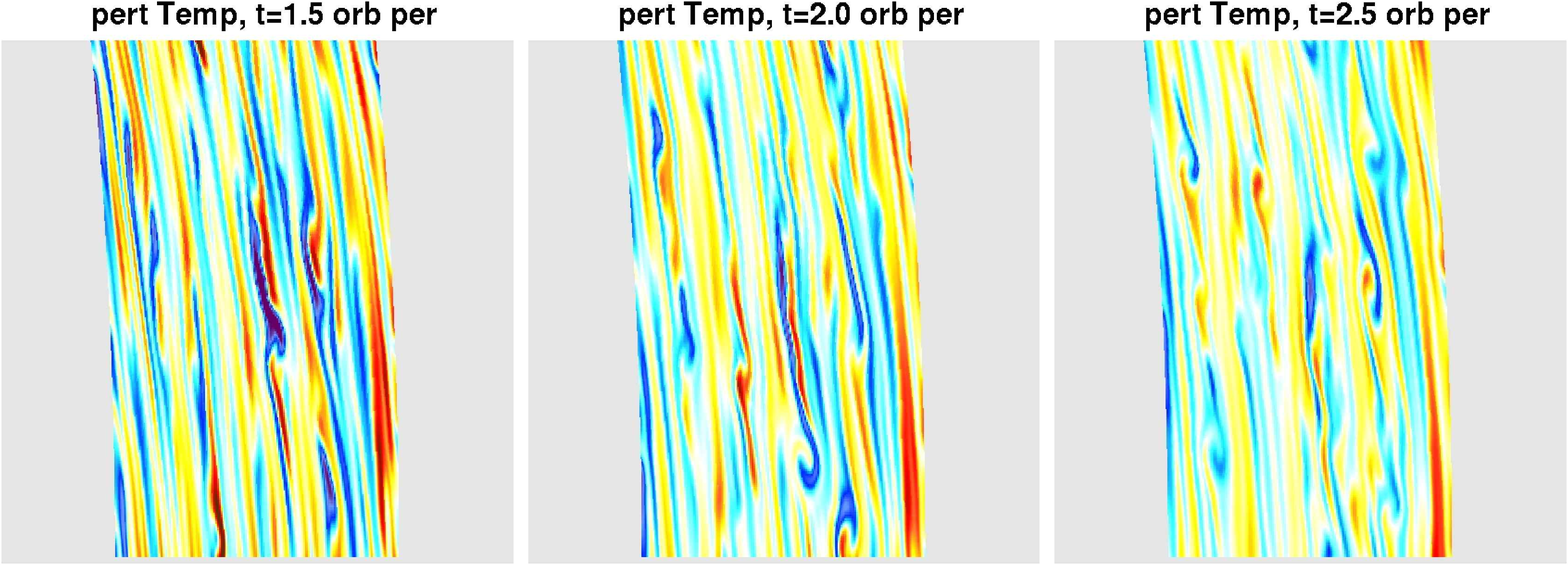}} 
\scalebox{.4}{\includegraphics{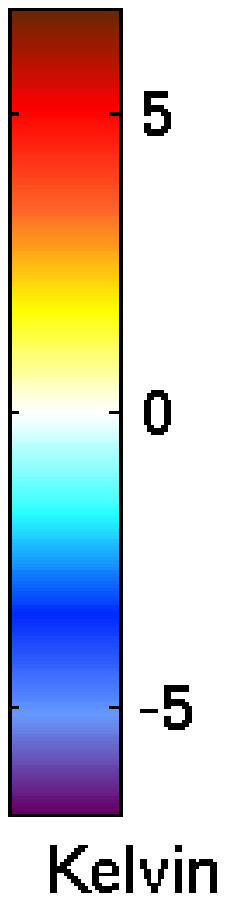}} 
 \caption{\label{f_early2} 
Labels as in Fig. \ref{f_early1}, but for later time.  After the initial shearing period, thin strips of vorticity roll up to form anticyclonic vortices, while the temperature field continues to decay.}
\end{figure}

\begin{figure}[tbh]
 \center
 \scalebox{.12}{\includegraphics{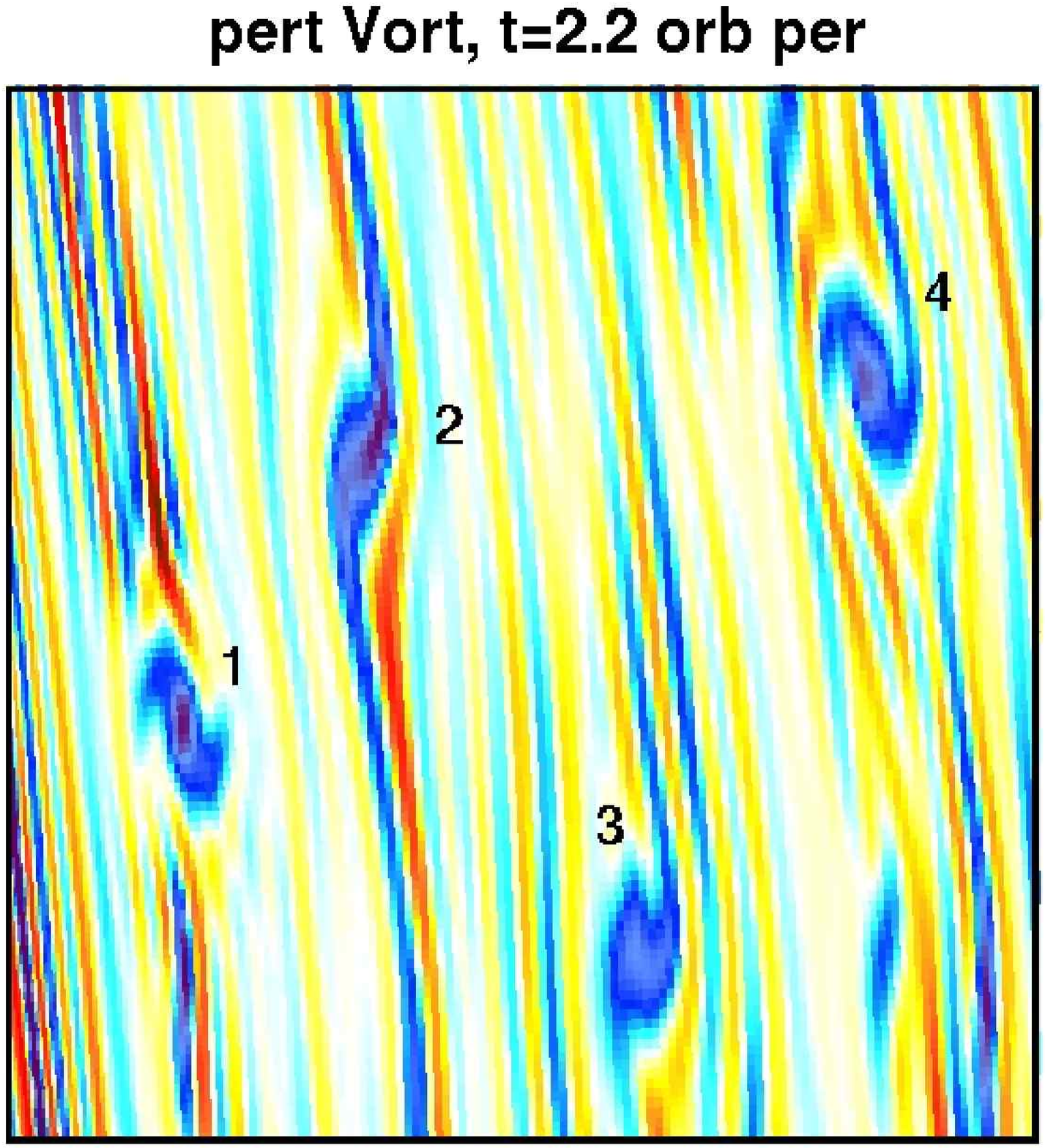}} 
 \scalebox{.4}{\includegraphics{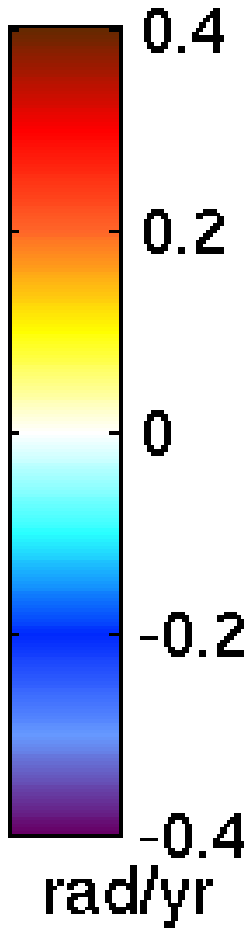}}\\ 
\scalebox{.12}{\includegraphics{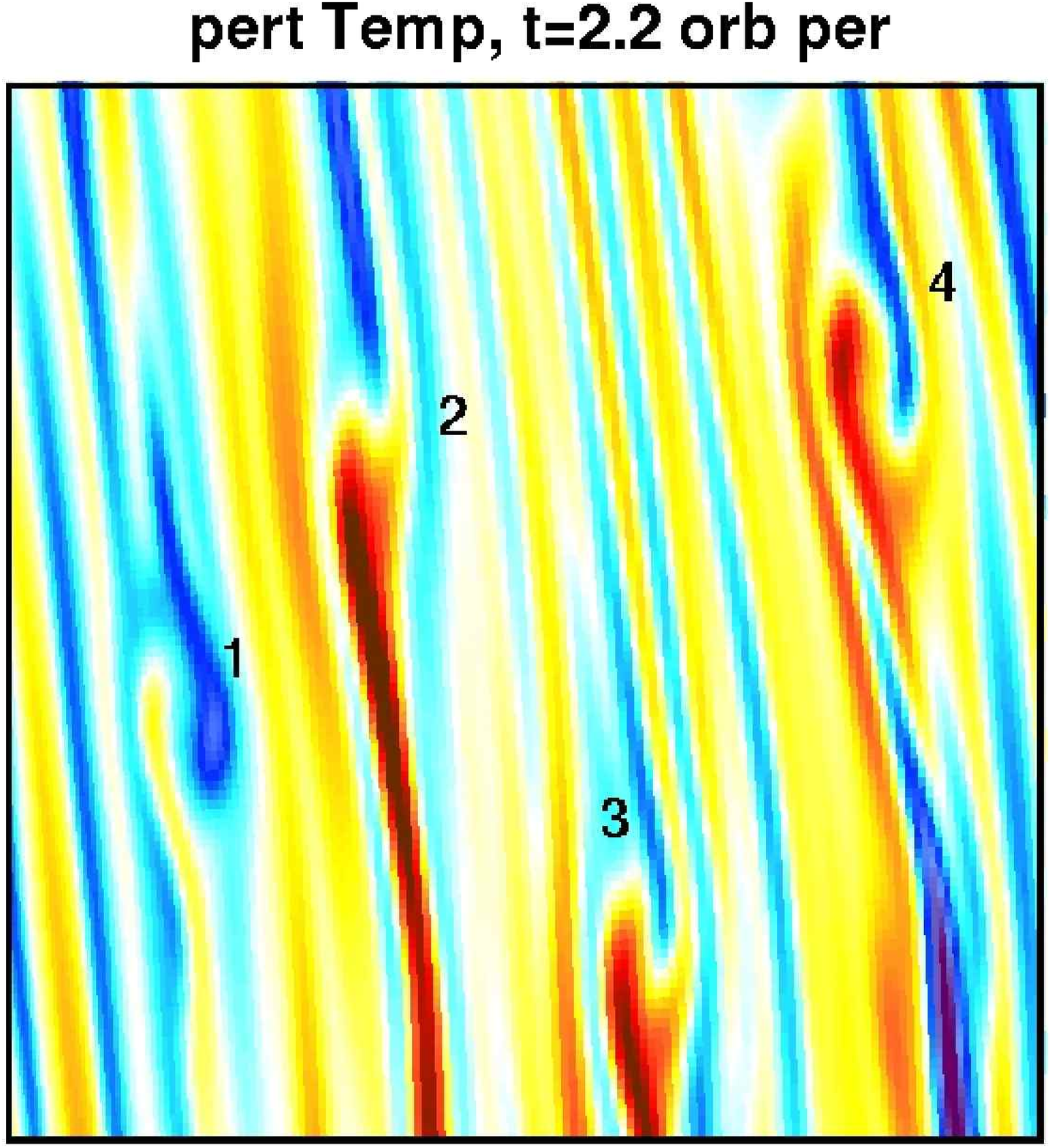}} 
\scalebox{.4}{\includegraphics{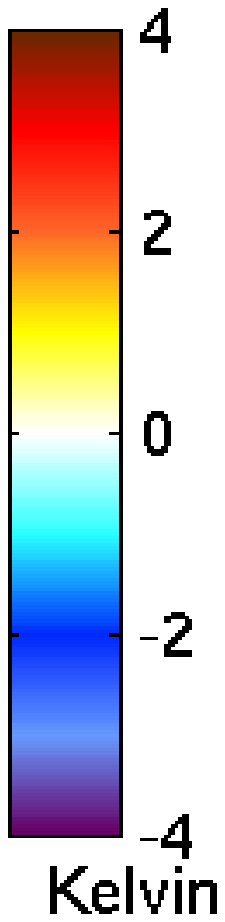}} 
 \caption{\label{f_zoom} 
Close-up view (0.4AU by 0.4 AU) of perturbation variables in simulation A.  Four anticyclones are labeled (top), with their corresponding temperature perturbations (bottom).  Each vortex advects warmer fluid inward (towards the left) and colder fluid outward, in a clockwise direction.}
\end{figure}

     Once  the radial profile of the total vorticity in the simulation develops
inflection points, it will satisfy the requirements for a secondary
instability \citep{Drazin_Reid81bk}.  In the simulation, this secondary
instability manifests itself after the first orbital period by causing regions
of negative vorticity adjacient to strong vorticity gradients patches to roll up
into anticyclonic vortices (vortices rotating in the opposite direction as the
background flow), while the positive vorticity regions tend to shear out and
eventually diffuse away (Fig. \ref{f_early2}).   Similar secondary instabilities have been
reported by \citet{Li_ea05apj} when simulating disks perturbed by an embedded
planet.  The persistence of anticyclones and the absence of cyclones is typical
of simulations with background differential rotation \citep{Godon_Livio99apj, Marcus90jfm,Marcus_ea00pp}.  The vortices are closely coupled to the
temperature perturbations.  Figure \ref{f_zoom} shows that each vortex advects cold fluid
from the outer annulus inwards and advects warm fluid from the inner annulus
outwards.  This net radial transport of heat down the background temperature
gradient plays an important roile in the baroclinic feedback and long-term
evolution of anticyclones, as discussed in Part II.


The baroclinic term is the only source of perturbation vorticity in the vorticity equation (\ref{num_vorticity}).  Since these simulations begin with zero vorticity perturbation, the baroclinic term must be responsible for the increase in vorticity at very early times.  To explore this further, simulation B repeated simulation A, but turned off the baroclinic term at $t=0.1,0.2,0.4,$ and $0.8$ orbital periods.  As soon as the baroclinic term is turned off, the vorticity and kinetic energy immediately drops (Fig. \ref{f_bcl_off_early}).  If the baroclinic term is turned off sufficiently early, no vortices form; the longer it stays on during the first orbital period, the stronger the vortices become.  These results show that the baroclinic term is responsible for the production of vorticity at early times.

\begin{figure}[tbh]
 \scalebox{.8}{\includegraphics{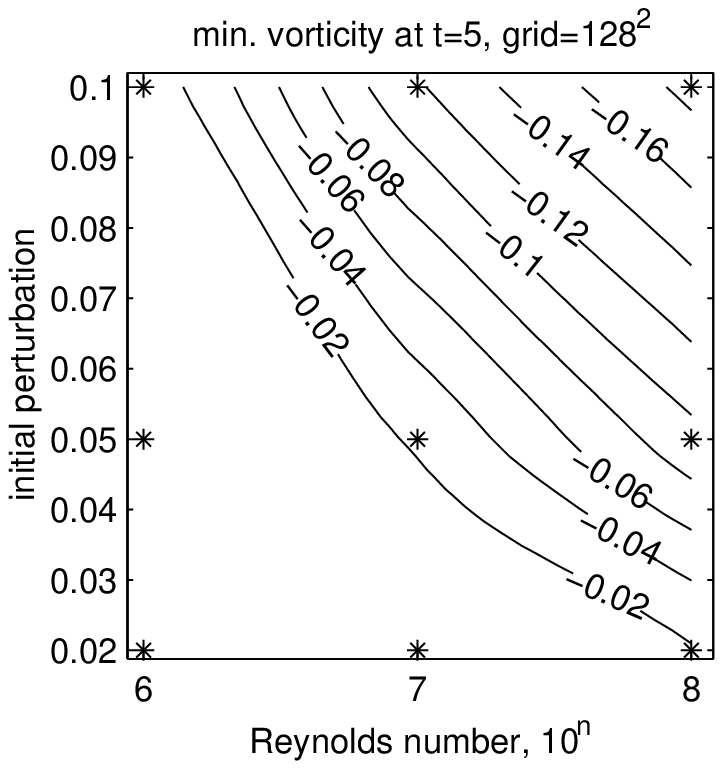}} 
 \scalebox{.8}{\includegraphics{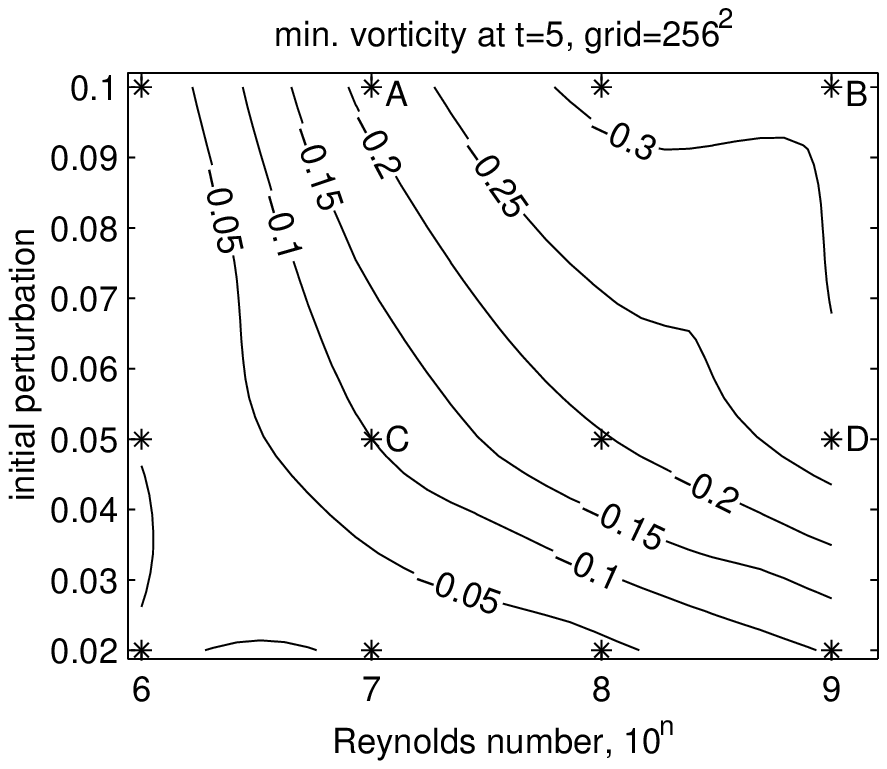}}\\  
\scalebox{.15}{\includegraphics{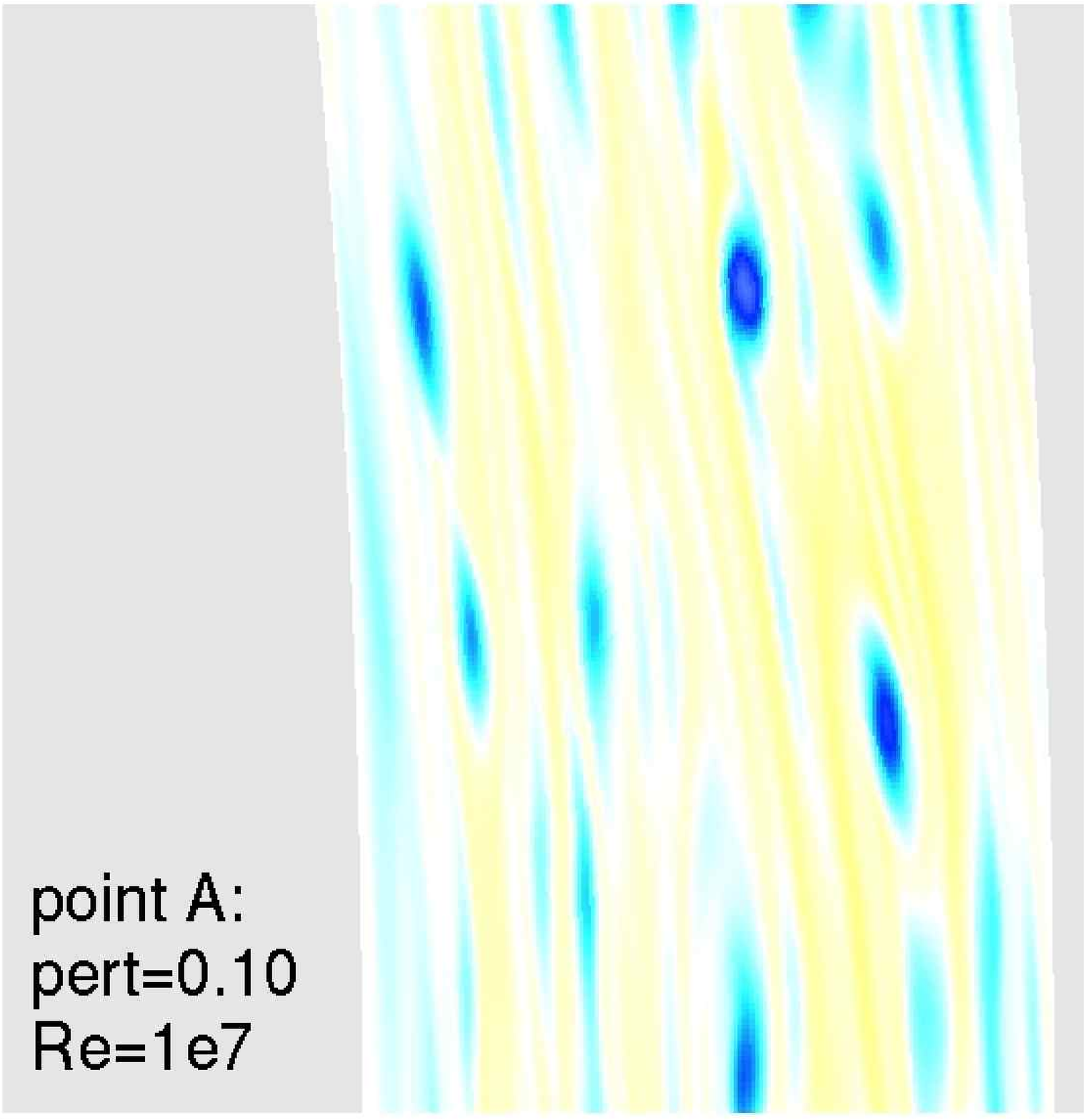}} 
\scalebox{.15}{\includegraphics{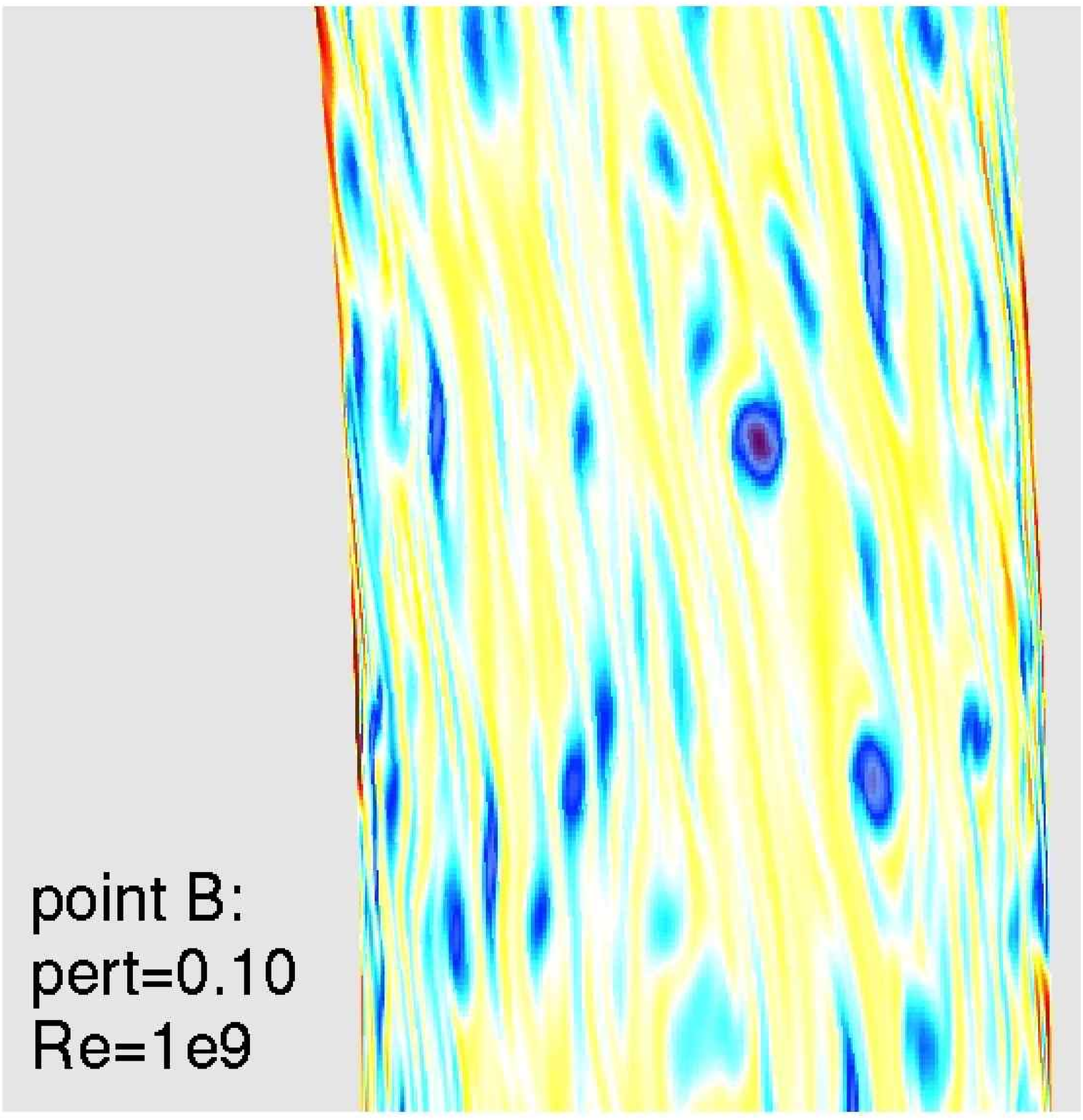}}\\  
\scalebox{.15}{\includegraphics{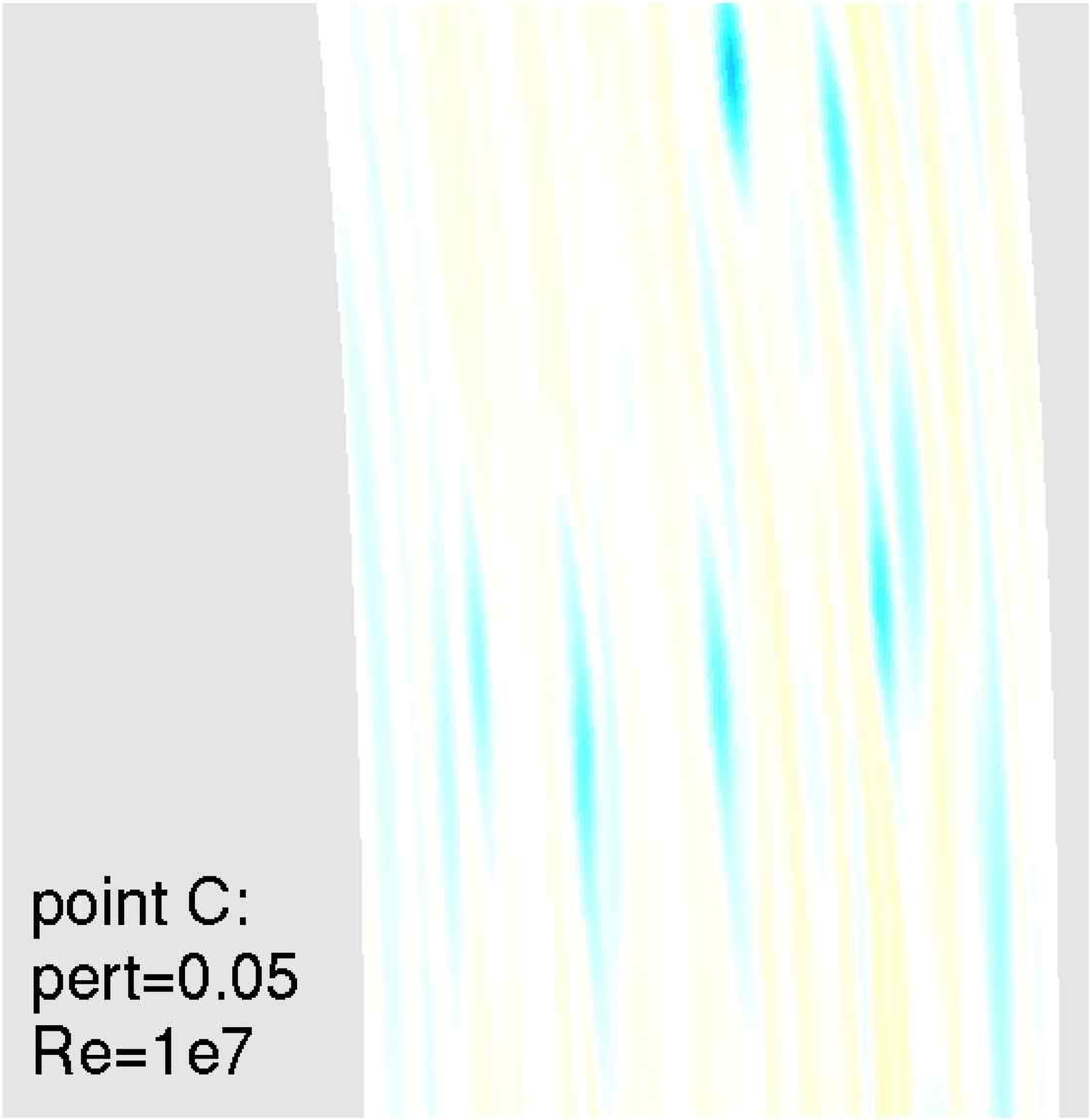}} 
\scalebox{.15}{\includegraphics{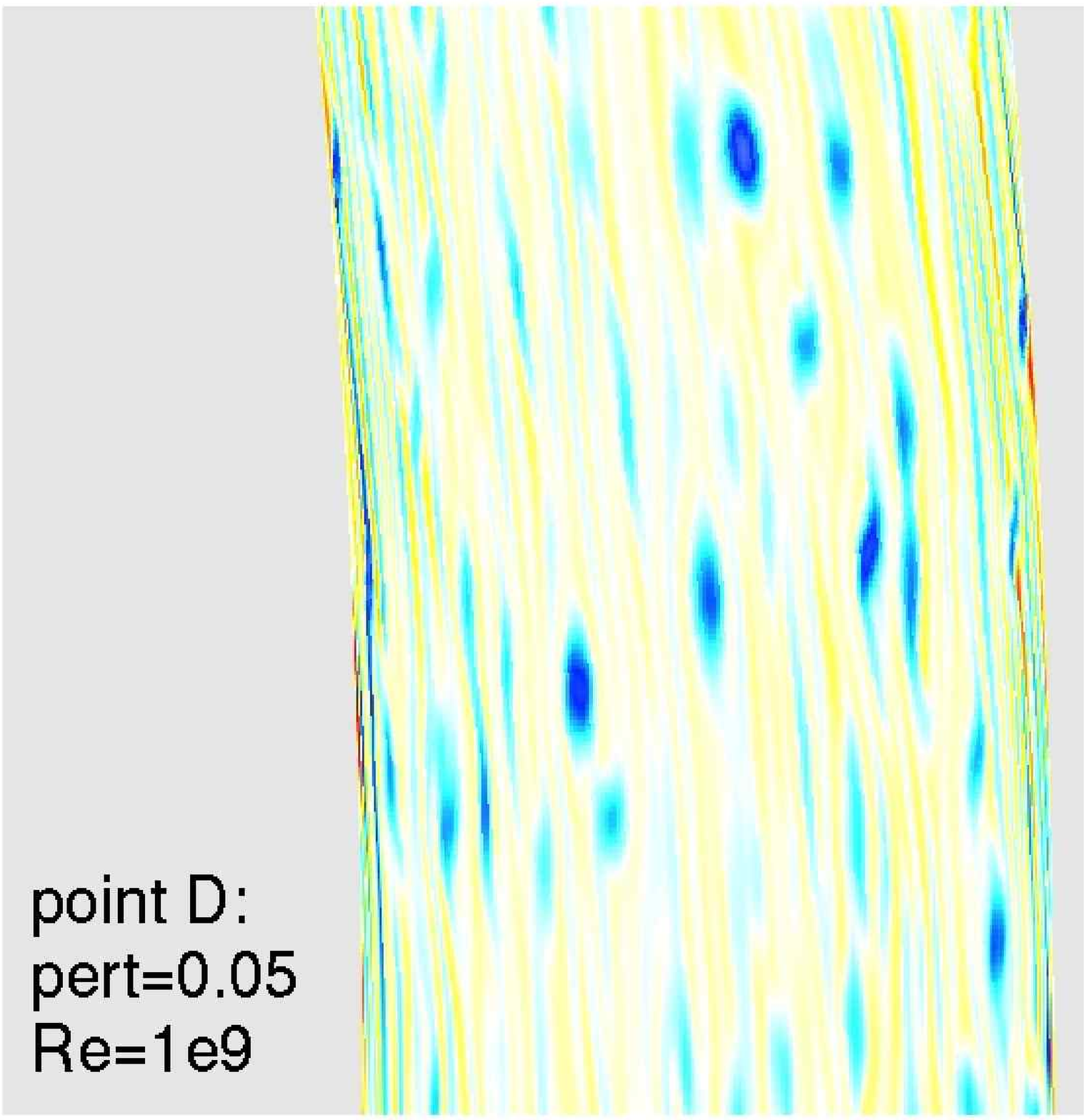}} 
\scalebox{.4}{\includegraphics{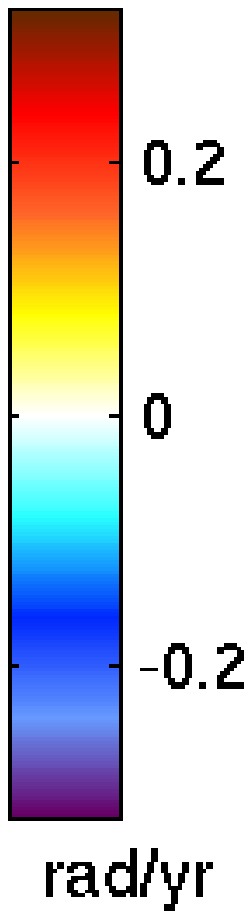}} 
 \caption{\label{f_sens_pert} 
Plots showing that vortex strength increases with either initial temperature perturbation, Reynolds number, or resolution.  The contour plots show the minimum vorticity after five orbital periods, as measured from simulation P with parameter values at each asterick.  These contours correspond to vortex strength, as shown in the four snapshots of perturbation vorticity $\zeta'$ at $t=5$, where points A to D are labelled on the contour plot.  
As simulations become more realistic with higher resolution and higher Reynolds number, a smaller initial perturbation is requried to initiate vortices.
}\end{figure}

\begin{figure}[tbh]
 \scalebox{.8}{\includegraphics{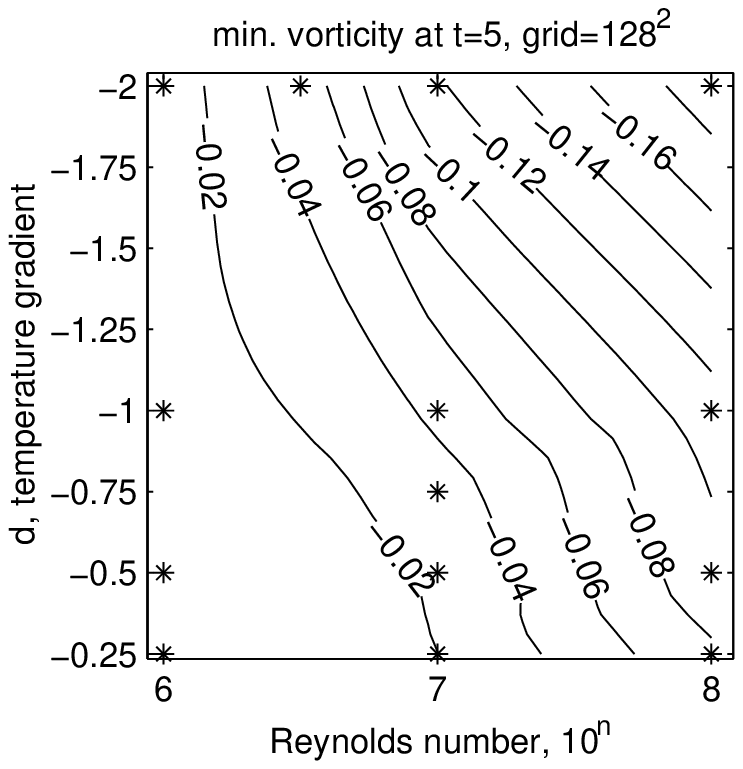}}  
 \scalebox{.8}{\includegraphics{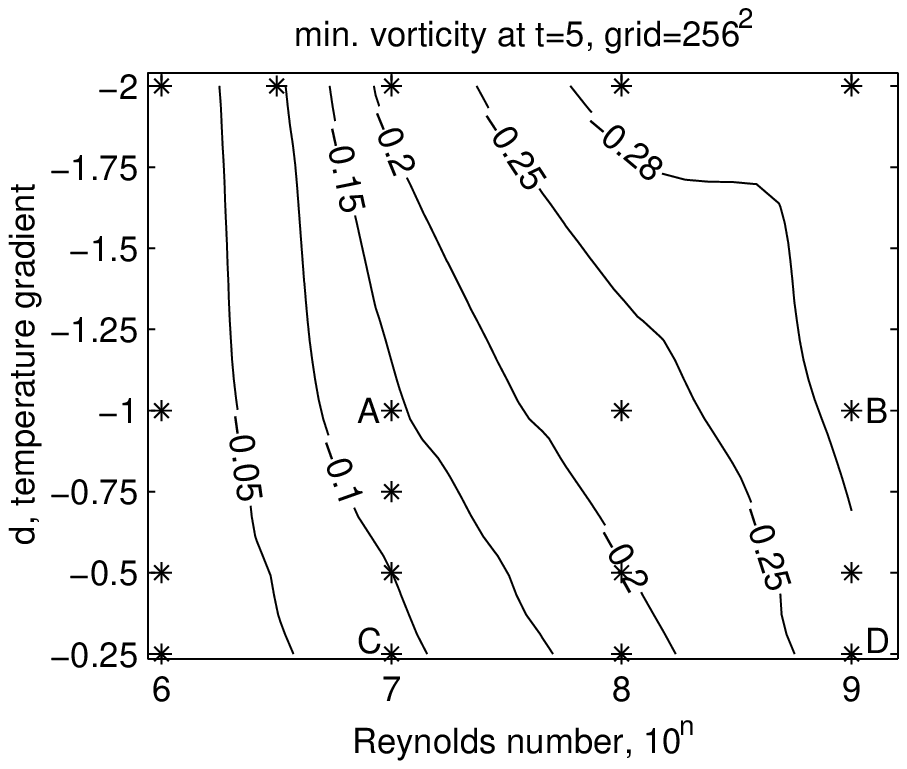}}\\  
\scalebox{.15}{\includegraphics{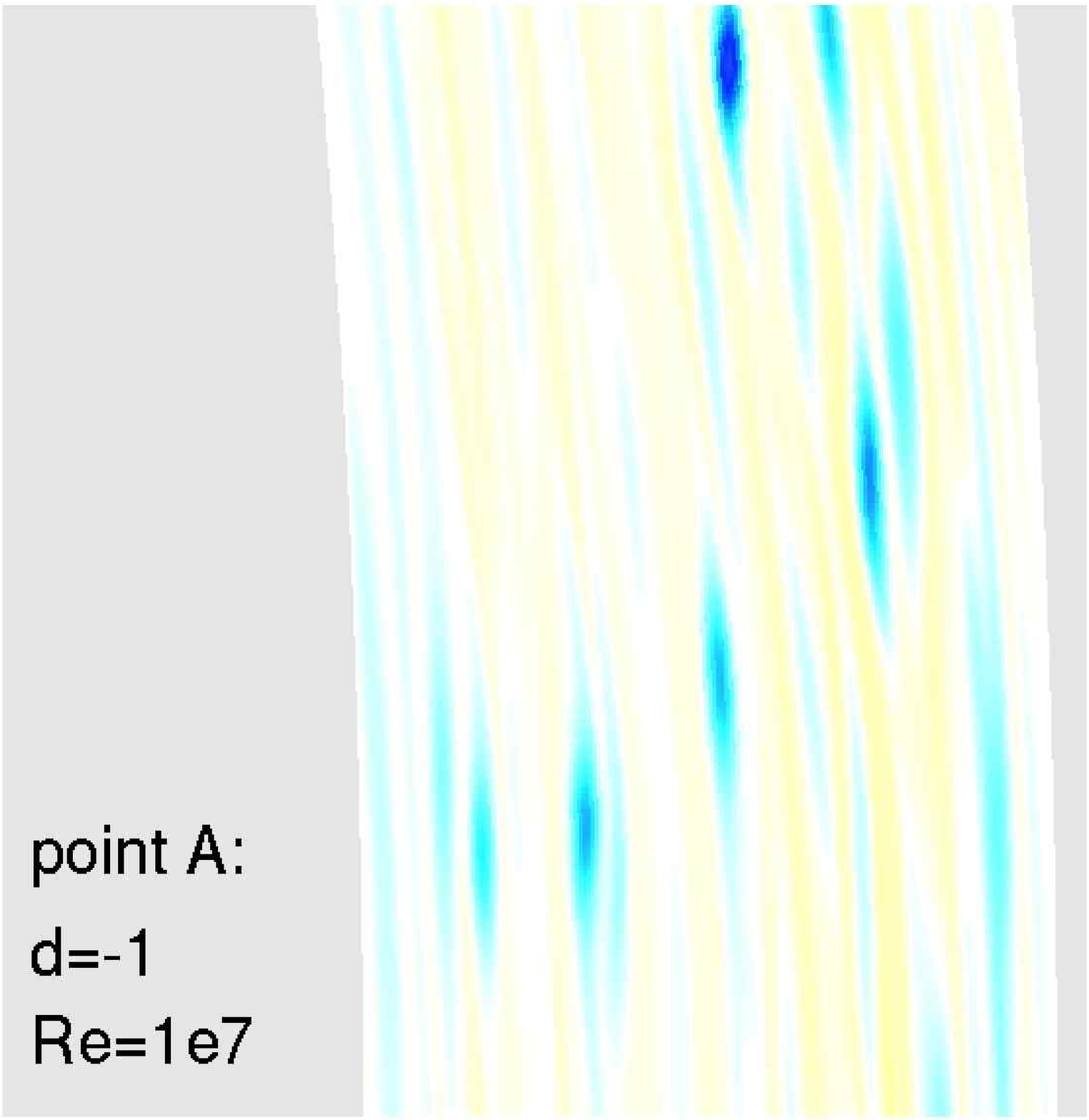}}  
\scalebox{.15}{\includegraphics{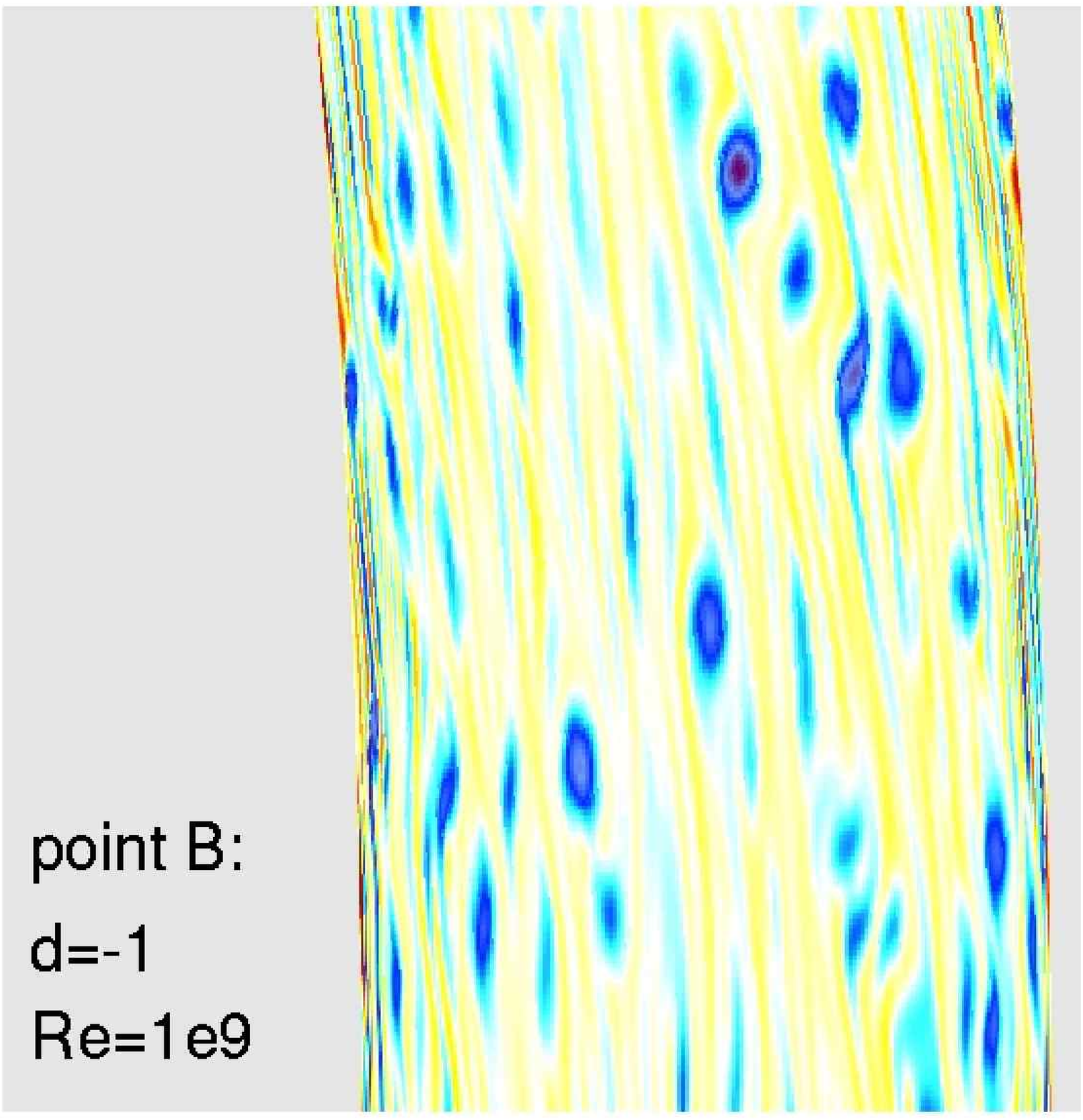}}\\  
\scalebox{.15}{\includegraphics{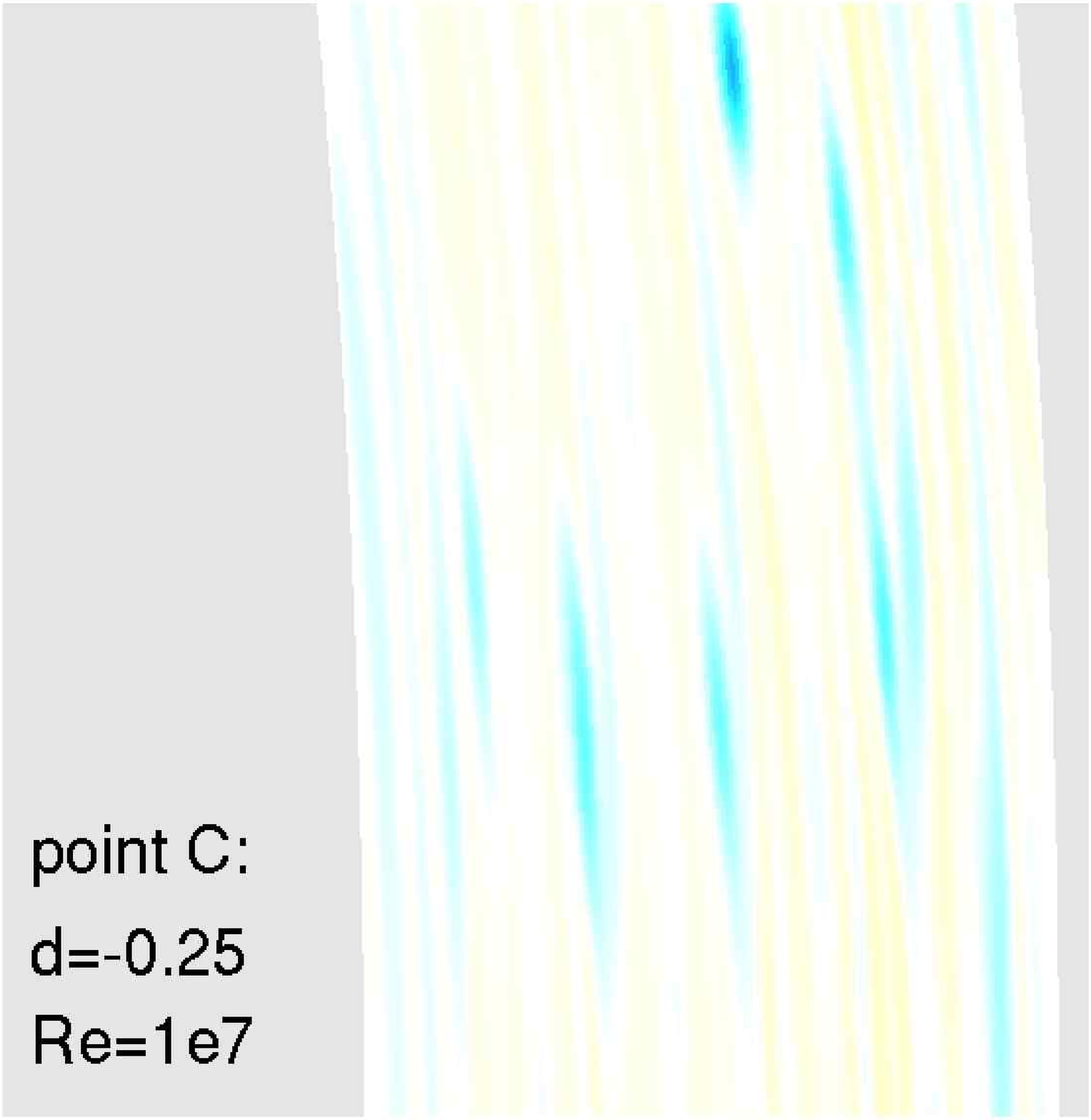}}  
\scalebox{.15}{\includegraphics{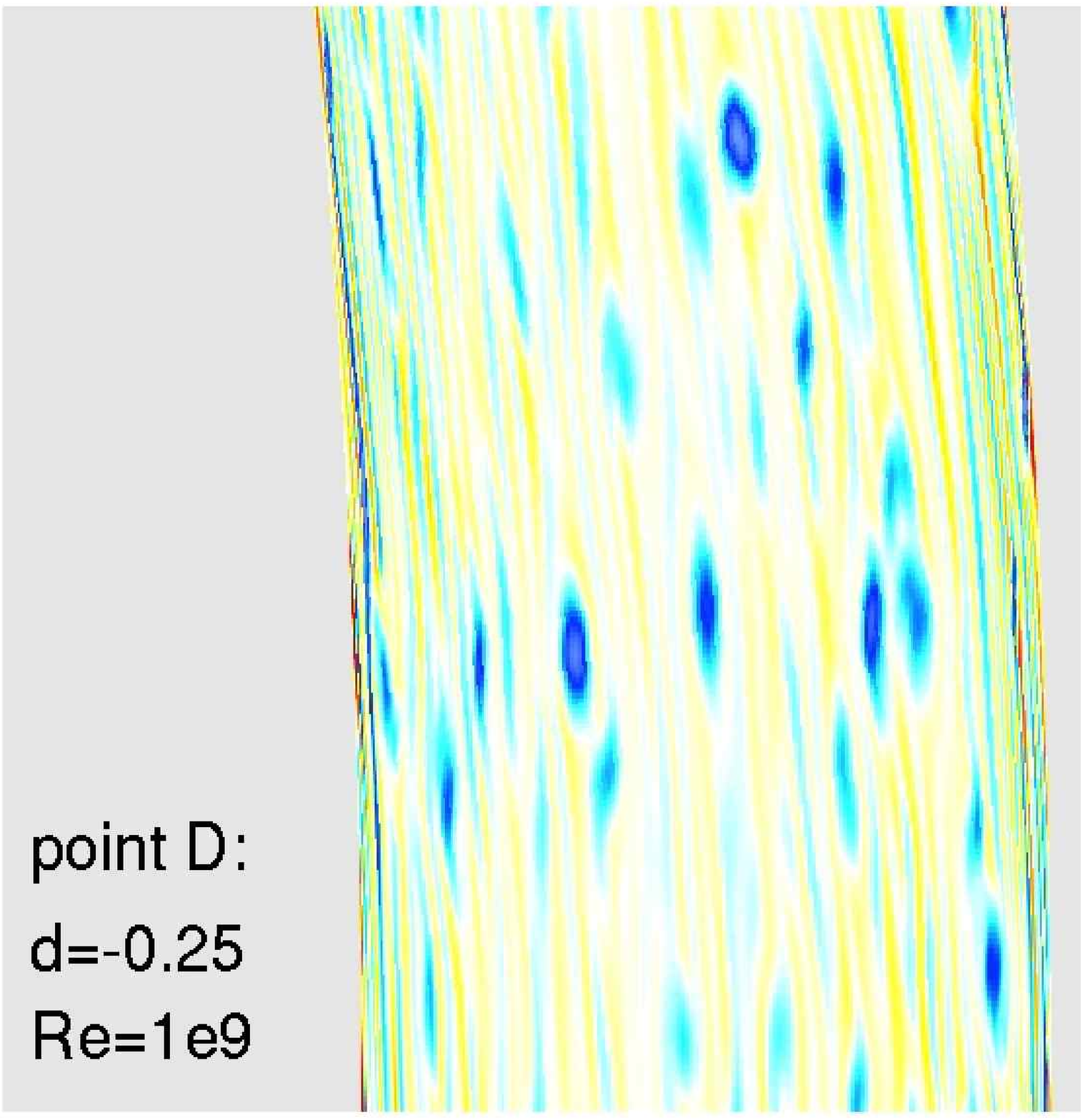}}  
\scalebox{.4}{\includegraphics{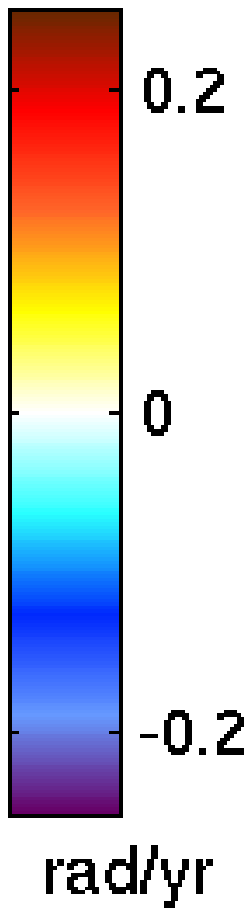}}  
 \caption{\label{f_sens_Tgrad} 
Plots showing that vortex strength increases with either the radial temperature gradient, Reynolds number, or resolution (simulation TG).  Plot description is the same as Fig. \ref{f_sens_pert}.  
This shows that higher resolution and higher Reynolds numbers allow less negative radial temperature gradients to initiate vortices. 
}\end{figure}

\begin{figure}[tbh]
 \scalebox{.8}{\includegraphics{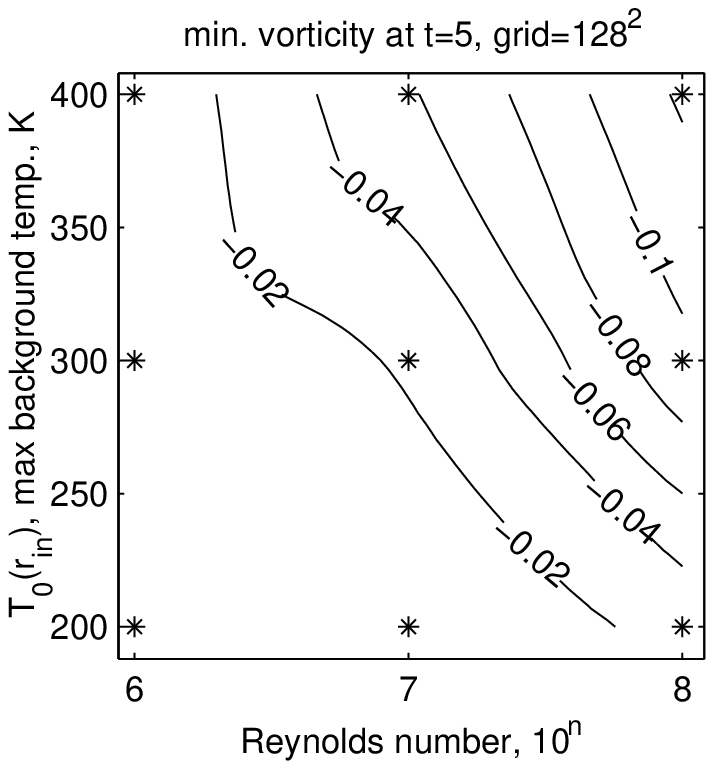}}  
 \scalebox{.8}{\includegraphics{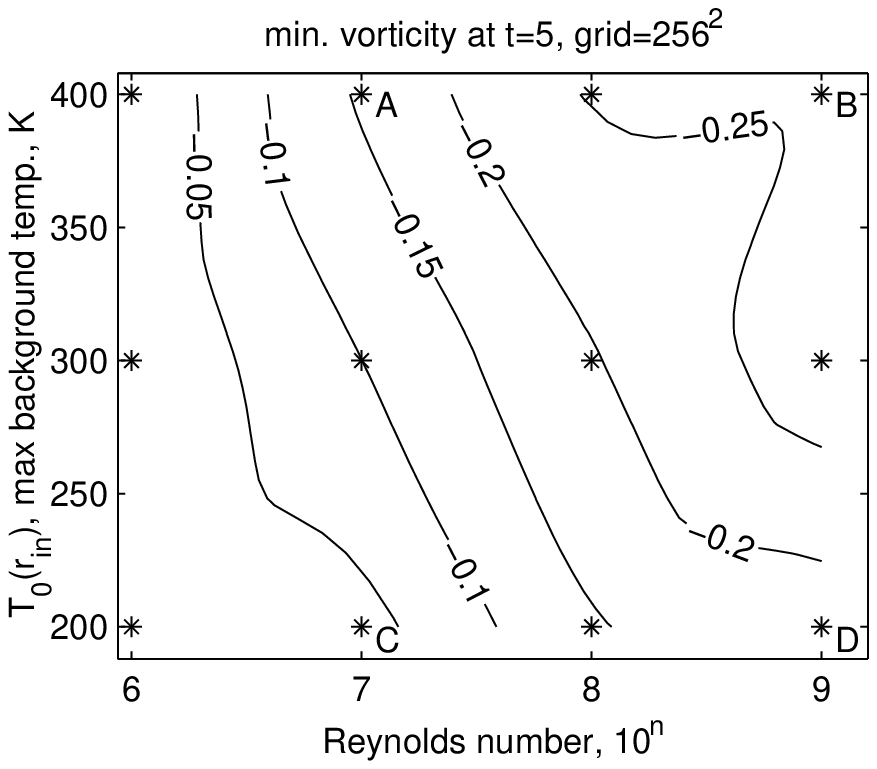}}\\  
\scalebox{.15}{\includegraphics{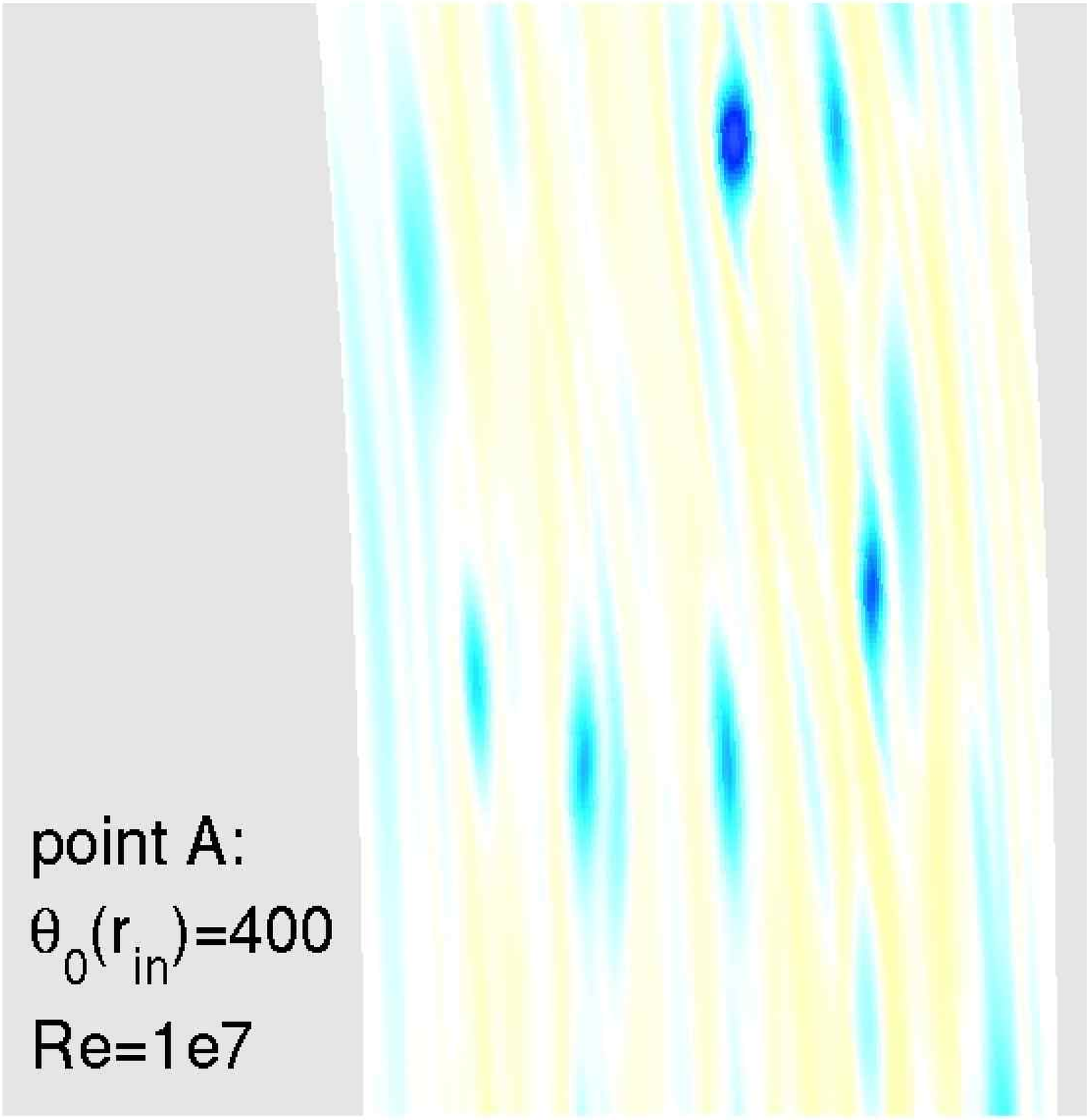}}  
\scalebox{.15}{\includegraphics{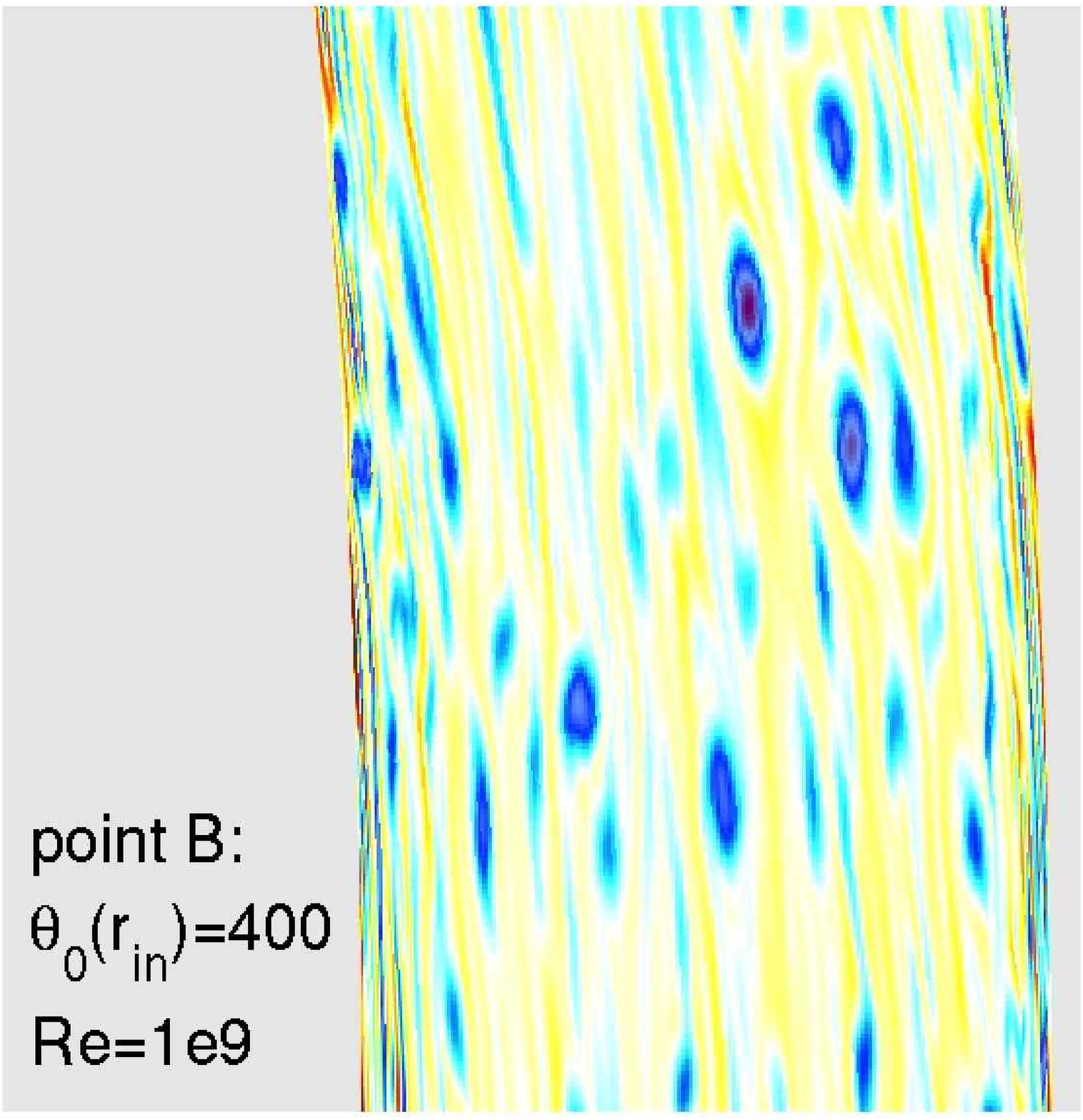}}\\  
\scalebox{.15}{\includegraphics{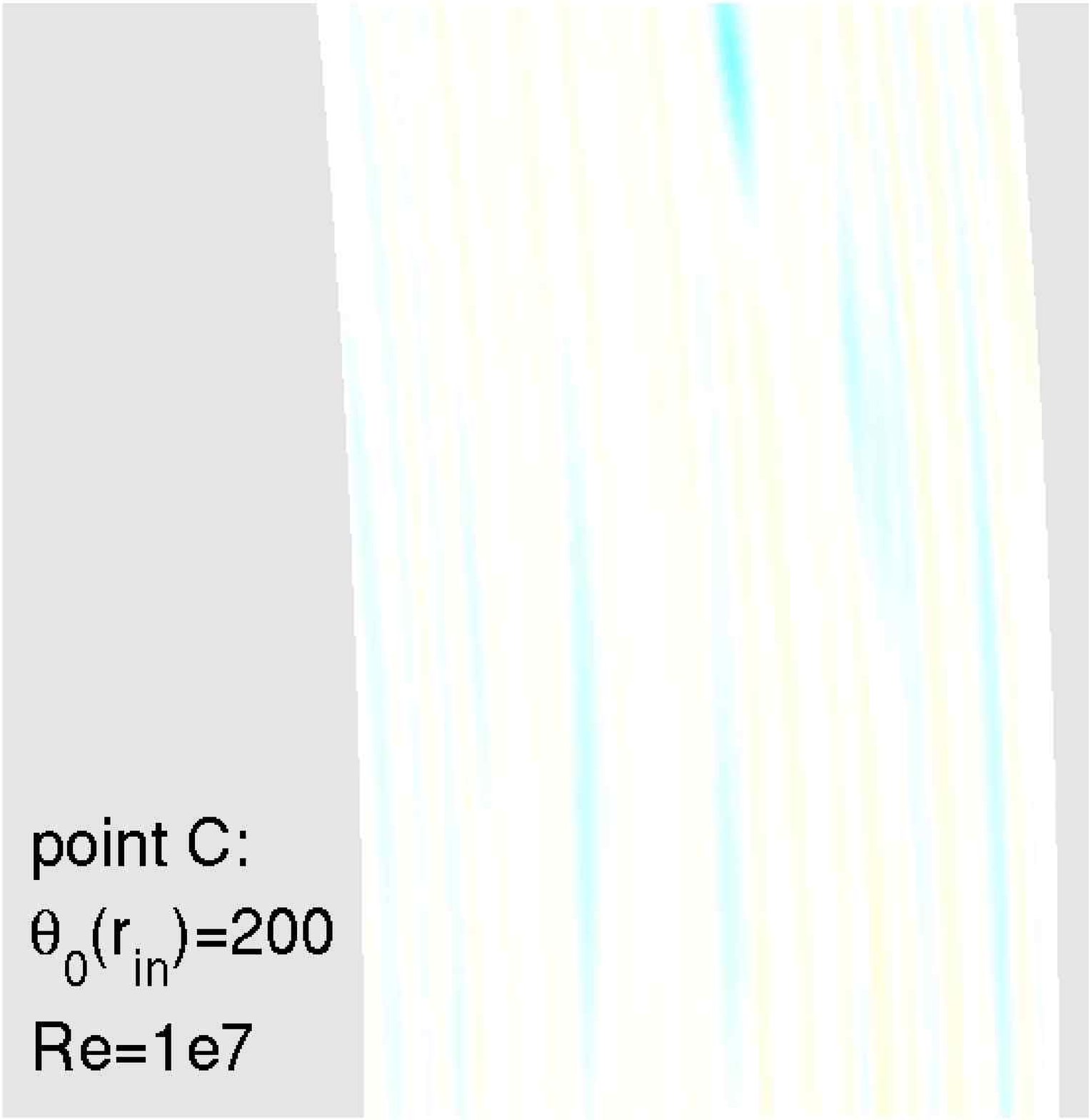}}  
\scalebox{.15}{\includegraphics{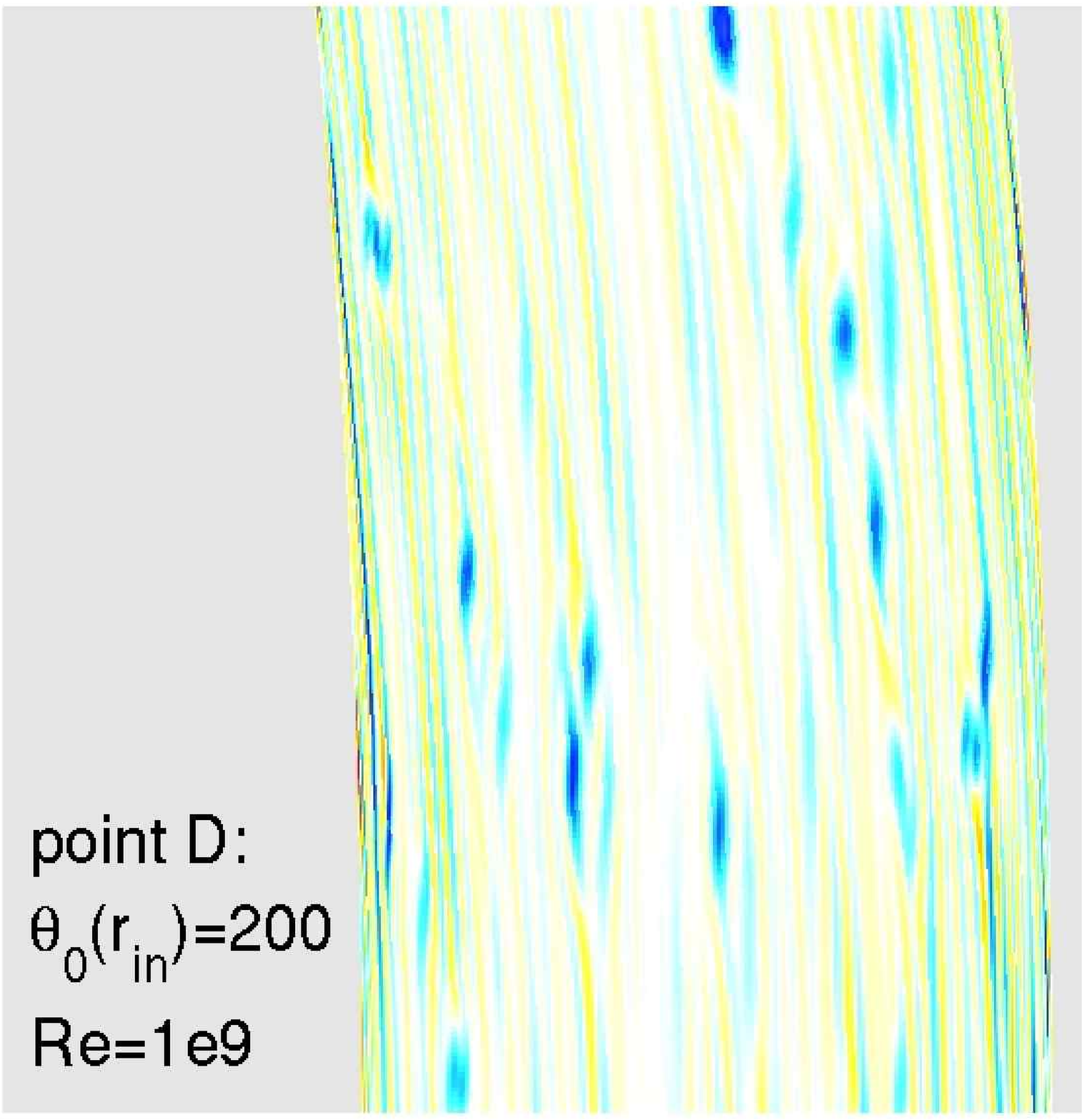}}  
\scalebox{.4}{\includegraphics{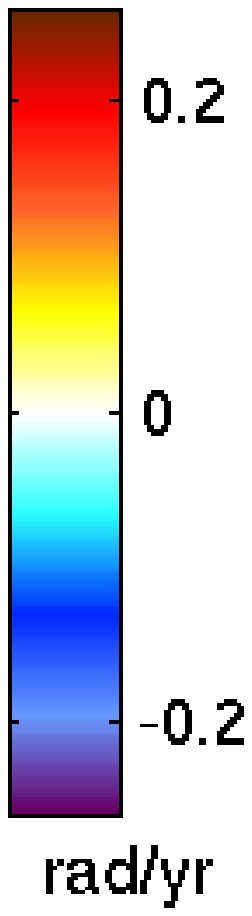}}  
 \caption{\label{f_sens_T} 
Plots showing that vortex strength increases with the background temperature, Reynolds number, or resolution (simulation T).  Plot description is the same as Fig. \ref{f_sens_pert}.    
}\end{figure}

The strength of vortex formation was found to depend on the following parameters: the background temperature gradient $d$, the magnitude of the background temperature $c$, the size of the initial temperature perturbation, the Reynolds number $Re$, and the resolution.  In order to explore this large parameter space, a sensitivity study consisting of 74 simulations was conducted.  These are simulations P, TG, and T in Table \ref{t_parameters}, and the results are shown in Figs. \ref{f_sens_pert}, \ref{f_sens_Tgrad}, and \ref{f_sens_T}, respectively.  The contour plots of minimum vorticity show the strength of the vortices after five orbital periods.  When all other parameters remain constant, stronger vortices result from higher Reynolds numbers and higher resolution.  These factors are not independent---higher resolution is required for higher Reynolds number.  At resolutions of $128\times 128$ and $256\times 256$ the highest attainable Reynolds numbers were $10^8$ and $10^9$, respectively.  These are still far from realistic Reynolds numbers for a protoplanetary disk, which would be $10^{12}$ or higher\footnote{\citet{Goldreich_Ward73apj} estimate the kinematic viscosity of a protoplanetary disk at 1AU to be $\nu=2\times10^6$ cm$^2$s$^{-1}$.  Using a length scale of 1AU and a time scale of one year, the Reynolds number $Re=L^2/(t\nu)=4\times10^{12}$.}.

The contours in Fig. \ref{f_sens_pert} slant from the top left to the bottom right of the contour plot.  This means that stronger vortices may be produced with either a stronger initial temperature perturbation or a higher Reynolds number.  Stated another way, as Reynolds number increases (i.e. becomes more realistic) a smaller initial perturbation is required to kick off vortex formation.  Thus even though we can produce vortices with a 5\% initial perturbation at $256\times 256$, the trend suggests that a much smaller perturbation would be required at higher resolutions and higher Reynolds numbers.  The contour plots in Figs. \ref{f_sens_Tgrad} and \ref{f_sens_T} show similar trends for the background temperature gradient $d$ and the magnitude of the background temperature $c$.  As shown by the snapshots of vorticity perturbation, when the Reynolds number is sufficiently high ($Re=10^9$), vortices form in these simulations when $T_0=300K(r/9.5AU)^{-0.25}$ (Fig. \ref{f_sens_Tgrad}, point D) and when $T_0=200K(r/9.5AU)^{-0.5}$ (Fig. \ref{f_sens_T}, point D).  In summary, vortex strength increases with: larger background temperature gradients (more negative $d$); warmer background temperatures $c$; larger initial temperature perturbations; higher Reynolds number; and higher resolution. 

     We can summarize the vorticity evolution displayed by our simulations with
the following steps:
(1) the initial potential temperature perturbations lead to baroclinic
production of vorticity perturbations; 
(2) the differential rotation of the background flow causes the vorticity
perturbations to be stretched out in the azimuthal direction and to develop large
radial gradients; 
(3) the radial gradients in vorticity perturbations become large enough to
trigger a secondary instability that is caused by local inflection points in the
radial profile of the total vorticity.   The secondary instability causes
regions of negative perturbation vorticity to roll up into nonlinear
anticyclones.
A major result of this paper is our finding that insufficient numerical
resolution or too large a numerical viscosity will cause the vorticity
perturbations to be damped out before step (3) can occur.  Conversely, our
simulations show that increasing both the numerical resolution and the effective
Reynolds number of our simulation allows weaker initial temperature
perturbations to evolve into nonlinear vortices.

\begin{figure}[tbh]
\center
\scalebox{.8}{\includegraphics{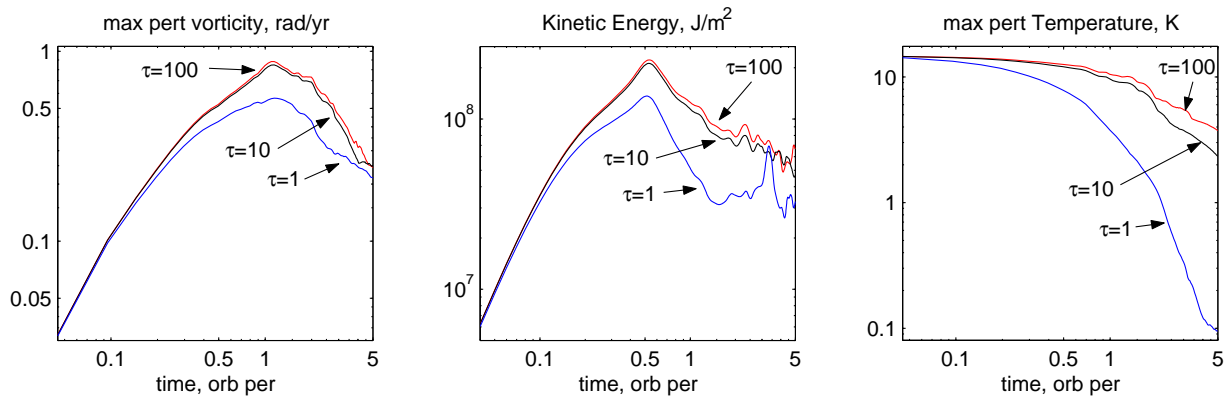}}  
\caption{\label{f_vary_tau_early} 
Data for simulation set Tau, where the radiative cooling rate is varied between $\tau=1$ and $\tau=100$ orbital periods.  With $\tau=1$ the disk cools faster (right), which weakens temperature gradients so that less vorticity is produced by the baroclinic feedback.  During vortex formation, vortices are strongest as $\tau\rightarrow\infty$, i.e., in the limit of no radiative cooling.
}\end{figure}

\begin{figure}[tbh]
\center
\scalebox{.8}{\includegraphics{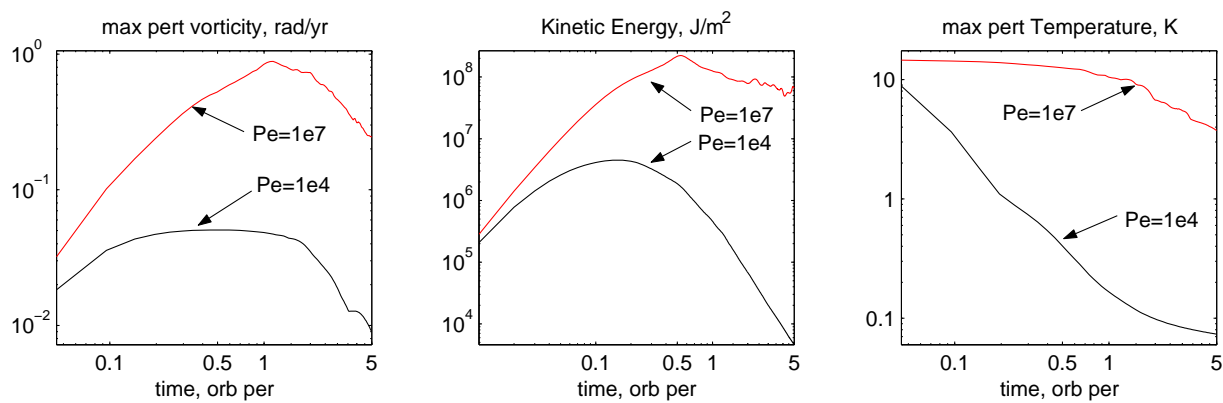}}  
\caption{\label{f_vary_Pe_early} 
Data for simulation set Pe.  When thermal diffusion is higher (lower Peclet number, $Pe=1e4$) the disk cools faster (right).  This weakens temperature gradients, making the baroclinic feedback weaker, so that growth in vorticity (left) and perturbation kinetic energy (center) is much slower.
}\end{figure}

In Part II the radiative cooling time $\tau$ and the Peclet number $Pe$ are shown to play an important role in the baroclinic feedback.  These parameters were varied in simulations Tau and Pe presented here to see their effects at early times.  The variables $\tau$ and $Pe$ control the rate of dissipation in the potential temperature equation (\ref{num_temperature}); small values of $\tau$ and $Pe$ (fast diffusion) return the perturbation potential temperature $\theta'$ to zero faster than large values.  This is exactly what is observed when either $\tau$ or $Pe$ are varied (Figs. \ref{f_vary_tau_early} and \ref{f_vary_Pe_early}).  As the temperature perturbations drop off, temperature gradients become smaller, and the baroclinic term produces less vorticity.  Thus at early times, faster dissipation (smaller $\tau$ or $Pe$) results in lower perturbation temperature, vorticity, and kinetic energy.  At early times, vortex formation is strongest in the limit of no thermal dissipation ($\tau,Pe\rightarrow\infty$).   Part II will show that faster dissipation has the opposite effect during the vortex growth phase; smaller values of $\tau$ or $Pe$ result in stronger vortices due to the baroclinic feedback.

\section{Conclusions \label{s_conclusions}}

In this paper we have: introduced a two-dimensional anelastic equation set which is a simplified model of a protoplanetary disk; described in detail a new numerical model of this equation set; and shown results of vortex formation in small-domain, short-time simulations.  The baroclinic term is responsible for the vortex production, as it is the only source term in the vorticity equation and the initial vorticity perturbation was zero.  Also, increases in vorticity and kinetic energy stop when the baroclinic term is turned off.  Each vortex has a characteristic temperature perturbation around it, due to the temperature advection about the vortex; this plays an important role in the baroclinic feedback described in Part II.  

Seventy-four simulations were conducted to test the sensitivity of vortex formation to several parameters.  It was found that vortex strength increases with: higher background temperature ($c$), stronger background temperature gradients (more negative $d$), larger initial temperature perturbations, higher Reynolds number, and higher resolution.  At the highest resolution and Reynolds number tested, $256\times 256$ and $Re=10^9$, vortices could be formed with a zero initial vorticity perturbation, and an initial temperature perturbation which was 5\% of the background, background temperatures of 200--300K, and background temperature gradients that vary radially as $r^{-0.25}$ and $r^{-0.5}$.

Additional simulations showed that in this early stage of vortex formation, increasing the thermal diffusion weakens vortex formation because the temperature perturbations die away too quickly.  In Part II we will see that faster thermal diffusion enhances the baroclinic feedback and makes vortices grow more quickly, once they have formed.

The equation set used in this study is anelastic and two-dimensional; this was chosen because it is the simplest model which can still include baroclinic effects and radiative cooling in a differentially rotating disk.  This simple equation set allowed us to create a fast and efficient numerical model that was then used to explore a large swath of parameter space with hundreds of simulations.  By choosing an anelastic 2D model, we limit ourselves to dynamics that can occur in that regime, and are blind to processes which may amplify or destabilize the vortices that we observe.  For example, the anelastic assumption means that the disk's perturbation velocities are subsonic.  The scale analysis in \citet{Barranco_Marcus05apj} showed that the horizontal extent of subsonic, compact vortices in a Keplerian shear cannot be much greater than the scale height of the disk.  Then they performed numerical simulations of these compact vortices;  when the vortices extend vertically through the disk they are unstable to small perturbations and are ripped apart by Keplerian shear.  In fact, only three-dimensional vortices that are centered above or below the midplane are robust in the long-term.

The vortices in our anelastic simulations are subsonic, compact (typically 0.005--0.1AU), and would be a vertical column of vorticity if they were transplanted from their 2D universe into a 3D universe.  These are like the vortices that \citet{Barranco_Marcus05apj} found to be unstable in 3D.  However, their model had a constant background temperature; our vortices require a sufficiently steep radial temperature gradient and do not develop with a constant background temperature.  So it seems that the vortices in \citet{Barranco_Marcus05apj} are not due to a baroclinic feedback, even though their model includes baroclinic effects.  Regardless of the formation process, the work of \citet{Barranco_Marcus05apj} suggests that columnar vortices in a 3D protoplanetary disk are unstable;  this provides a strong motivation for a follow-up of our study using 3D simulations.  It is possible that the parameters that influence the baroclinic formation and growth of vortices in 2D (like ours) would affect short, stable, off-midplane 3D vortices (like those in \citealt{Barranco_Marcus05apj}) in the same manner.  The current 2D study is important because it elucidates the effects of various parameters on the baroclinic instability in an idealized 2D simulation.  It also allows one to estimate the numerical resolution and effective Reynolds number that would be required in a 3D simulation to see the instability.  

By describing the equation set, numerical model, and sensitivity study of parameters, this paper lays the groundwork for Part II, which studies the role of the baroclinic feedback in growth and longevity of vortices.  The simulations in Part II use a quarter annulus domain, which is ten times larger than the small domain used here, making it effectively a lower resolution model.  Thus the findings of the sensitivity analysis apply---in order to form vortices, a higher background temperature, steeper background temperature gradient, and larger initial perturbation were required.  Because of the sensitivity analysis of Part I, it is clear that all of these requirements ease with higher resolution and Reynolds number.

Finally, it is useful to compare model parameters to those inferred from observations of real protoplanetary disks.  \citet{Beckwith_ea90aj} surveyed 86 pre-main-sequence stars in the Taurus-Aurign dark clouds, and found that disk temperatures range from 80--300K and vary radially as $r^d$ with $d\le-0.5$.  Our simulations produced vortices with background temperatures of 200--300K when $d=-0.5$, indicating that vortices in our model can be generated under realistic conditions.

\section{Acknowledgements}
We thank P. Marcus for insightful feedback and practical advice,
A.P. Boss for useful discussions, and an anonymous referee for criticism which significantly improved the final version.
MRP has been supported by an NSF Vigre Grant,
DMS-9810751, awarded to the Applied Mathematics Department at the
University of Colorado at Boulder.  KJ has been supported by NSF grant
OCE-0137347 as well as the University of Colorado Faculty Fellowship. 
GRS was supported by NASA's Origins of Solar Systems research program.

\appendix
\section{Details of the Numerical Model\label{a_model_details}}
\setcounter{equation}{0} 
\renewcommand{\theequation}{A\arabic{equation}}
\newcommand{\R}{\hat{R}}

\subsection{Time-stepping scheme \label{a_time_stepping}}
The third order, three stage Runge-Kutta scheme by \citet{Spalart_ea91jcp} is used to solve the model equations (\ref{num_stream_fxn}-\ref{num_temperature}).  The two Jacobian terms are nonlinear and will be treated explicitly, while the linear terms on the right will be treated implicitly.  

Define the Jacobians 
$M(\Psi,\zeta) =  \p\left(\Psi,\zeta/\Sigma_0\right)$ and
$N(\Psi,\theta) = \p\left(\Psi,\theta\right)/\Sigma_0$
so that each stage of the Runge-Kutta routine is
\bea
&&  \left(1+\beta \Delta t \tau^{-1} 
         - \beta \Delta t \kappa_e\del^2\right) \theta'^n
= \left(1-\alpha\Delta t\tau^{-1} 
         + \alpha\Delta t \kappa_e\del^2\right) \theta'^{n-1}
\nn \\ && \hspace{1cm}
   - \gamma \Delta t N^{n-1} - \delta \Delta t N^{n-2} \label{runge-kutta_T}
\\ 
&&  \left(1- \beta \Delta t \nu_e \del^2\right) \zeta'^n
= \left(1+ \alpha\Delta t \nu_e \del^2\right) \zeta'^{n-1}
\nn \\ && \hspace{1cm}
  - \gamma \Delta t M^{n-1} - \delta \Delta t M^{n-2}   
    + \frac{\Delta t c_p}{r}\frac{\p \pi_0}{\p r} 
    \left(\beta\frac{\p \theta'^n}{\p\phi}
      +\alpha\frac{\p \theta'^{n-1}}{\p\phi}\right) 
\label{runge-kutta_q}
\eea
where $\alpha,\beta,\gamma,\delta$ are constants of the scheme, $n$ is the current stage, and $\Delta t$ is the time step (other variables are defined in Sect. \ref{s_cuppd_equations}).  At the beginning of this stage, all $n-1$ and $n-2$ variables are known. (At stage 1 the $n-2$ variables are zero).  Then $\theta'^n$ and $\zeta'^n$ are calculated by inverting operators of the form $(a+b\nabla^2)$ in spectral space (Sect. \ref{a_inversion}).  The perturbation stream function $\Psi'^n$ is found by inverting the Laplacian-like operator in (\ref{num_stream_fxn}) in spectral space (Sect. \ref{a_stream_function}).  Finally, $M^n$ and $N^n$ are found by transforming derivatives of $\theta'^n$, $\zeta'^n$, and $\Psi'^n$ into physical space, computing the Jacobians, and transforming back into spectral space.

\subsection{Inverting $\left(a+b\nabla^2\right)$ \label{a_inversion}}

Consider the general problem 
\bea
\left(a+b\del^2\right) U(r,\phi) = R(r,\phi),
\eea
which in polar coordinates is
\bea
\left(a+ b\frac{\p^2}{\p r^2} + \frac{b}{r}\frac{\p}{\p r}
 +  \frac{b}{r^2}\frac{\p^2}{\p \phi^2}\right)U = R. \label{polar_laplacian1}
\eea
The radial basis functions are Chebyshev polynomials, which have a natural domain of $z\in[-1,1]$.  To transform from an annular domain with $r\in[r_1,r_2]$, define $f=(r_2-r_1)/2$ and $g=(r_1+r_2)/2$ so that $r=fz+g$.  By the chain rule 
$\p_r U=\p_z U/f$ so that (\ref{polar_laplacian1}) is now 
\bea
\left( a+ \frac{b}{f^2}\frac{\p^2}{\p z^2} 
 + \frac{b}{f(fz+g)}\frac{\p}{\p z}
 +  \frac{b}{(fz+g)^2}\frac{\p^2}{\p \phi^2}\right)U &=& R.
\label{polar_laplacian2}
\eea

In order to invert this operator in spectral space, expand the variables as
\bea
U(z,\phi) = \sum_{p=0}^P\sum_{k=0}^K u_{p,k}T_p(z)e^{ik\phi}
\eea
where $T_p$ is the $p^{th}$ Chebyshev mode.  Azimuthal derivatives are simply $\p_{\phi}^2=-k^2$, so that each Fourier mode is independent.  The radial derivatives are much more complicated; the products of $z$ and derivatives of $z$ which appear in (\ref{polar_laplacian2}) can be represented by summations with higher modes (\citealt{gottlieb_orszag77bk} p. 159, \citealt{Mason_Handscomb02bk}, p. 32) so that (\ref{polar_laplacian2}) can be written as $\mathbf{Au}_k=\mathbf{Br}_k$ where $\mathbf{A}$ is an upper triangular matrix, $\mathbf{B}$ is a pentadiagonal matrix, and $\mathbf{u}_k$ and $\mathbf{r}_k$ hold the Chebyshev coefficients of $U$ and $R$ for each Fourier mode $k$.  Because of the particular form of $\mathbf{A}$, we may add and subtract the rows of $\mathbf{Au}_k=\mathbf{Br}_k$  in a clever way to reduce $\mathbf{A}$ to a banded diagonal of width six.  Full details and lengthy algebraic manipulations can be found in \cite{Petersen04thesis}.

\subsection{The stream function\label{a_stream_function}}
After the vorticity $\zeta'^n$ is found using (\ref{runge-kutta_q}), the stream function $\Psi'^n$ is found by inverting the Laplacian-like operator in (\ref{num_stream_fxn}) in spectral space.  This inversion is difficult if the surface density $\Sigma_0(r)$ is an arbitrary function of $r$, but simplified if it is a power function $\Sigma_0=ar^b$.  Then  (\ref{num_stream_fxn}) can be rewritten as 
\bea
\ds   \left(\frac{\p^2}{\p r^2} 
   +  \frac{1-d}{r} \frac{\p }{\p r}
   + \frac{1}{r^2 } \frac{\p^2}{\p \phi^2} \right) \Psi'^n
= \Sigma_0(r) \zeta'^n
\eea
which is nearly the same form as (\ref{polar_laplacian1}).  The product $\Sigma_0(r) \zeta'^n$ must be computed in physical space and then transformed to spectral space for the inversion.

\subsection{Boundary conditions - the tau method\label{a_bc}}
Within spectral methods, there are three methods of satisfying the boundary conditions: the pseudospectral, Galerkin, and tau methods \cite[p 162]{Fornberg96bk}.  In the pseudospectral method, basis functions are chosen that each naturally conform to the boundary conditions.  In our problem, the azimuthal direction employs Fourier basis functions, which are naturally periodic.  In the Galerkin method, a new set of basis functions which satisfy the boundary condition are created from linear combinations of the old basis functions.  

In the tau method, which we use in the radial, the two highest mode equations in $\mathbf{Au}_k=\mathbf{Br}_k$ are replaced by equations which enforce the boundary conditions.  For Dirichlet boundary conditions $U(r_1,\phi)=C(\phi)$ and $U(r_2,\phi)=D(\phi)$ the corresponding tau lines are
\bea 
\sum_{p=0}^{P}u_{n,p}T_p(-1) = c_k, \;\;\;\;\;
\sum_{p=0}^{P}u_{n,p}T_p(1) = d_k , \;\;k=0\ldots K
\eea
and for Neumann boundary conditions $\p_rU(r_1,\phi)=C(\phi)$ and $\p_rU(r_2,\phi)=D(\phi)$ they are
\bea 
\sum_{p=0}^{P}u_{n,p}\p_z T_p(-1) = c_k, \;\;\;\;\;
\sum_{p=0}^{P}u_{n,p}\p_z T_p(1) = d_k, \;\;k=0\ldots K,
\eea
where $c_k$ and $d_k$ are the Fourier coefficients of $C(\phi)$ and $D(\phi)$.  In practice $C$ and $D$ are constants, so all modes are zero except $c_0$ and $d_0$.  Dirichlet boundary conditions are used for impermeable, stress-free boundaries, and Neumann are used for zero thermal flux boundaries.

\subsection{Stress-free boundary \label{a_stress_free_bc}}

The impermeable, stress-free radial boundary is enforced using the influence matrix technique \citep{Peyret02bk}.  Stress-free boundaries in polar coordinates are complicated by the form of the stress tensor so that the conditions for vorticity and stream function are coupled.  The momentum stream function is defined as $\Sigma_0 {\bf u} = -\del \times \Psi'$, so that the radial and azimuthal background velocities are
$u' = -\p_\phi\Psi'/r \Sigma_0$ and $v' = -\p_r\Psi'/\Sigma_0$, and the vorticity is defined in (\ref{num_vorticity}).  An impenetrable boundary implies that $u'=0$, which can easily be satisfied by requiring that $\Psi'$ be a constant along the boundary.  Stress-free boundaries mean that each component of the stress tensor
\bea
\sigma = \mu\left[
\ba{cc}
2\ds\frac{\p u'}{\p r} & 
  \ds\frac{1}{r}\frac{\p u'}{\p \phi} + \frac{\p v'}{\p r} - \frac{v'}{r} \\ \\
  \ds\frac{1}{r}\frac{\p u'}{\p \phi} + \frac{\p v'}{\p r} - \frac{v'}{r} &
     \ds\frac{2}{r}\frac{\p v'}{\p \phi} + \frac{2u'}{r}
\ea \right]
\eea
is zero (\citet{Aris62bk}, p. 181).  A stream function which is constant along the boundary in $\phi$ implies that $\p_\phi \Psi=0$, so that $u'=0$, $\p_r u'=0$, $\p_\phi u'=0$ and $\p_\phi v' =0$ on the boundary.  The remaining condition in the stress tensor is
\bea
\frac{\p v'}{\p r} - \frac{v'}{r}=0.  \label{stress_bc_v}
\eea

The stress-free impermeable boundary conditions in terms of stream function and vorticity are
\bea
&\Psi'=0, \\
& \ds
  \zeta' - \frac{2}{r\Sigma_0} \frac{\p \Psi'}{\p r} = 0 \label{stress_bc_q}
\eea
where the second condition comes from (\ref{stress_bc_v}).  These coupled equations are more complicated than Dirichlet or Neumann boundary conditions described in the previous section.  They require the influence matrix technique, where each unknown is written as the sum of a particular and homogeneous solutions
\bea
\left\{\ba{rcl}
\Psi' &=& \Psi^p + \alpha\Psi^- + \beta\Psi^+\\
\zeta' &=& \zeta^p + \alpha\zeta^- + \beta\zeta^+
\ea \right. \label{full_solutions}
\eea
where $\alpha$ and $\beta$ are unknowns to be solved for, superscript $p$ denotes the particular solution and superscripts $+$ and $-$ denote the homogeneous solutions.  A linear combination of two homogeneous solutions is required to construct the desired values at the inner and outer boundary of the annulus.  To that end, let $r^-$ and $r^+$ be the radial location of the inner and outer boundary, and require $\zeta^-(r^-)=1$ and $\zeta^+(r^+)=1$.  These are the only functions which contribute the the value of $\zeta$ on the boundary, so set $\zeta^-(r^+)=0$, $\zeta^+(r^-)=0$, and $\zeta^p(r^\pm)=0$.  The stream function $\Psi$ should always be zero at the boundary, so require that $\Psi^p(r^\pm)=0$ and $\Psi^\pm(r^\pm)=0$.

Boundary conditions are enforced using the Tau method during the inversion of Laplacian-type operators.  The vorticity is solved for in each stage of Runge-Kutta time-stepping using (\ref{runge-kutta_q}), which can be written as $(a+b\del^2)\zeta' = R$ where $R$ is the right-hand side.  The stream function is computed from the vorticity using (\ref{num_stream_fxn}), which will be abbreviated here as $\del^2_\Sigma \Psi = \zeta$.  The particular solutions must satisfy
\bea
(a+b\del^2)\zeta^p = R, \;\;\; \del^2_\Sigma \Psi^p = \zeta^p
\eea
while the homogeneous solutions must satisfy 
\bea
(a+b\del^2)\zeta^\pm = 0, \;\;\; \del^2_\Sigma \Psi^\pm = 0,
\eea
so that the full solutions (\ref{full_solutions}) satisfy 
\bea
(a+b\del^2)\zeta' = R, \;\;\; \del^2_\Sigma \Psi' = \zeta'.
\eea

The only condition which remains to be enforced is the stress-free boundary condition (\ref{stress_bc_q}), which is what caused all of this trouble in the first place.  It can be rewritten in terms of the particular and homogeneous solutions (\ref{full_solutions}) as
\bea
\left( \zeta^- - \frac{2}{r\Sigma_0}\frac{\p\Psi^-}{\p r} \right)\alpha
+ \left( \zeta^+ - \frac{2}{r\Sigma_0}\frac{\p\Psi^+}{\p r} \right)\beta
=     -\zeta^p + \frac{2}{r\Sigma_0}\frac{\p\Psi^p}{\p r}.
\eea
This must be enforced at the inner and outer boundaries, $r^-$ and $r^+$, resulting in the $2\times 2$ system of equations with unknowns $\alpha$ and $\beta$.  The final vorticity and stream function are then constructed with $\alpha$ and $\beta$ as in (\ref{full_solutions}).

In practice this influence matrix technique is performed for each Fourier mode $k$, so that (\ref{full_solutions}) is now 
\bea
\left\{\ba{rcl}
\Psi'_k &=& \Psi_k^p + \alpha_k\Psi_k^- + \beta_k\Psi_k^+\\
\zeta'_k &=& \zeta_k^p + \alpha_k\zeta_k^- + \beta_k\zeta_k^+
\ea \right. \label{full_solutions_k}
\eea
The full procedure conducted at every Runge-Kutta stage is
\begin{enumerate}
\item Solve $(a+b\del^2)\zeta^p = R$ and $\del^2_\Sigma \Psi^p = \zeta^p$ for $\zeta^p$ and $\Psi^p$.
\item Solve $(a+b\del^2)\zeta^\pm = 0$ and $\del^2_\Sigma \Psi^\pm = \zeta^\pm$ for $\zeta^\pm$ and $\Psi^\pm$.
\item For each Fourier mode $k$, solve
\bea \ba{rcl}
\left.\left( \ds\zeta_k^- - \frac{2}{r\Sigma_0}\frac{\p\Psi_k^-}{\p r} \right)
   \right|_{r^-} \alpha_k
+ \left.\left(\ds\zeta_k^+ - \frac{2}{r\Sigma_0}\frac{\p\Psi_k^+}{\p r} \right)
   \right|_{r^-}\beta_k
&=&\left.\left(\ds-\zeta_k^p + \frac{2}{r\Sigma_0}\frac{\p\Psi_k^p}{\p r}\right)
   \right|_{r^-} \\
\left.\left(\ds \zeta_k^- - \frac{2}{r\Sigma_0}\frac{\p\Psi_k^-}{\p r} \right)
   \right|_{r^+} \alpha_k
+ \left.\left(\ds \zeta_k^+ - \frac{2}{r\Sigma_0}\frac{\p\Psi_k^+}{\p r} \right)
   \right|_{r^+}\beta_k
&=& \left.\left(\ds -\zeta_k^p + \frac{2}{r\Sigma_0}\frac{\p\Psi_k^p}{\p r}\right)
   \right|_{r^+}
\ea \eea
for $\alpha_k$ and $\beta_k$.
\item Reconstruct the full solution using (\ref{full_solutions_k}).
\end{enumerate}


\newpage
\bibliographystyle{apj}
\bibliography{planetary_disk,cfd,qg,dynamics,updf_qg,my_pubs}

\clearpage

\clearpage

 \begin{table}[btp]
 \center
  \begin{tabular}[c]{c|ccccccc}
   name & $d$ & $c$ & init. pert. &$\tau$& grid & $Re$& Pe \\ 
   \hline
   A  &  -0.75 & 300 & 0.05 & 100 & $256^2$ & $10^8$& $10^7$ \\ 
   B  &  -0.75 & 300 & 0.05 & 100 & $256^2$ & $10^8$& $10^7$ \\ 
   TG  & {\bf -0.25-- -2}& 300 & 0.05 &1 &$128^2$, $256^2$&$10^6$--$10^9$& $10^7$ \\ 
   T  & -0.5 &{\bf 200--400} & 0.05  &1 &$128^2$, $256^2$&$10^6$--$10^9$ & $10^7$\\  
   P  & -0.5 & 300 & {\bf 0.02--0.1} &1 &$128^2$, $256^2$&$10^6$--$10^9$ & $10^7$ \\ 
   Tau  & -0.75 & 300 & 0.05 &{\bf 1--100} &$256^2$&$10^8$ & $10^7$ \\ 
   Pe  & -0.75 & 300 & 0.05 &100 &$256^2$&$10^8$ & {\bf $10^4,10^7$ } \\ 
 \end{tabular}
 \caption{\label{t_parameters} Model parameters for the numerical simulations discussed in this paper.  Here $d$ is the power on the background temperature function, $c$ is the background temperature at the inner radius, init.\ pert.\ is the magnitude of the initial temperature perturbation relative to the background, $\tau$ is the radiative cooling time, $Re$ is the Reynolds number, and $Pe$ is the Peclet number.  Simulation B is identical to A except that the baroclinic term is turned off during the simulation.
} \end{table}


\end{document}